\documentclass[twoside]{elsart}
\usepackage[dvips]{graphicx}
\usepackage{amssymb}

\newcommand{\pr}{\textit{Phys. Rev.\/}\ }
\newcommand{\prep}{\textit{Phys. Rep.\/}\ }
\newcommand{\pl}{\textit{Phys. Lett.\/}\ }
\newcommand{\prl}{\textit{Phys. Rev. Lett.\/}\ }
\newcommand{\np}{\textit{Nucl. Phys.\/}\ }
\newcommand{\vol}[3]{\textbf{#1}\textrm{(#2)#3}}
\newcommand{\tf}{\mbox{$\theta_{\mathrm{flow}}$}}
\newcommand{\etran}{\mbox{$E_{\mathrm{t}}$}}
\newcommand{\etranlcp}{\mbox{$E_{\mathrm{t12}}$}} 
\newcommand{\fox}{\mbox{$H_2$}}
\newcommand{\riso}{\mbox{$R_{\mathrm{iso}}$}}

\newcommand{\eiso}{\mbox{$E_{\mathrm{iso}}$}}

\newcommand{\mutot}{\mbox{$N_{\mathrm{C}}$}}
\newcommand{\nf}{\mbox{$N_{\mathrm{f}}$}}
\newcommand{\mimf}{\mbox{$N_{\mathrm{IMF}}$}}
\newcommand{\mlcp}{\mbox{$N_{\mathrm{LCP}}$}}
\newcommand{\ztot}{\mbox{$Z_{\mathrm{tot}}$}}

\renewcommand{\deg}{\mbox{$^o$}}
\newcommand{\mpn}{\mbox{A.MeV}}
\newcommand{\fpi}{\mbox{$4 \pi$}}
\newcommand{\gu}{\nuc{155}{Gd} + \nuc{nat}{U}}

\newcommand{\gc}{\nuc{155}{Gd} + \nuc{12}{C}}
\newcommand{\gdu}{\nuc{155}{Gd} + \nuc{nat}{U} 36~A.MeV}

\newcommand{\twid}{\mbox{$\sim$}}

\newcommand{\upto}{\mbox{---}}

\newcommand{\enex}{\mbox{$E^*$}}

\begin{document}

\begin{frontmatter}
\title{
Multifragmentation of a very heavy nuclear
system (I): Selection of single-source events\thanksref{Ganil}}
\thanks[Ganil]{Experiment performed at Ganil}

\author[ipno]{J.D. Frankland\thanksref{pres-ganil}},
\author[ipno]{Ch.O.~Bacri},
\author[ipno]{B.~Borderie},
\author[ipno]{M.F.~Rivet},
\author[ipno]{M.~Squalli},
\author[ganil]{G.~Auger},
\author[lpc]{N.~Bellaize},
\author[lpc]{F.~Bocage},
\author[lpc]{R.~Bougault},
\author[lpc]{R.~Brou},
\author[cea]{Ph.~Buchet},
\author[ganil]{A.~Chbihi},
\author[lpc]{J.~Colin},
\author[lpc]{D.~Cussol},
\author[cea]{R.~Dayras},
\author[ipnl]{A.~Demeyer},
\author[cea]{D.~Dor\'e},
\author[lpc]{D.~Durand},
\author[ipno,cnam]{E.~Galichet},
\author[lpc]{E.~Genouin-Duhamel},
\author[ipnl]{E.~Gerlic},
\author[ipnl]{D.~Guinet},
\author[ipnl]{Ph.~Lautesse},
\author[ganil]{J.L.~Laville},
\author[lpc]{J.F.~Lecolley},
\author[cea]{R.~Legrain},
\author[lpc]{N.~Le Neindre},
\author[lpc]{O.~Lopez},
\author[lpc]{M.~Louvel},
\author[ipnl]{A.M.~Maskay},
\author[cea]{L.~Nalpas},
\author[lpc]{A.D.~Nguyen},
\author[buch]{M.~Parlog},
\author[lpc]{J.~P\'{e}ter},
\author[ipno]{E.~Plagnol},
\author[napol]{E.~Rosato},
\author[ganil]{F.~Saint-Laurent\thanksref{cada}},
\author[ganil]{S.~Salou},
\author[lpc]{J.C.~Steckmeyer},
\author[ipnl]{M.~Stern},
\author[buch]{G.~T\u{a}b\u{a}caru},
\author[lpc]{B.~Tamain},
\author[ganil]{O.~Tirel},
\author[ipno]{L.~Tassan-Got}, 
\author[lpc]{E.~Vient}, 
\author[cea]{C.~Volant},
\author[ganil]{J.P.~Wieleczko}

\collaboration{INDRA collaboration}

\address[ipno]{Institut de Physique Nucl\'eaire, IN2P3-CNRS, F-91406 Orsay
 Cedex, France.}
\address[ganil]{GANIL, CEA et IN2P3-CNRS, B.P.~5027, F-14076 Caen Cedex, France.}
\address[lpc]{LPC, IN2P3-CNRS, ISMRA et Universit\'e, F-14050 Caen Cedex,
France.}
\address[cea]{DAPNIA/SPhN, CEA/Saclay, F-91191 Gif sur Yvette Cedex,
France.}
\address[ipnl]{Institut de Physique Nucl\'eaire, IN2P3-CNRS et Universit\'e,
F-69622 Villeurbanne
Cedex, France.}
\address[cnam]{Conservatoire Nationale des Arts et
M\'etiers, 292 rue Saint-Martin, 75141 Paris Cedex 03, France.}
\address[buch]{National Institute for Physics and Nuclear Engineering,
RO-76900 Bucharest-M\u{a}gurele, Romania}
\address[napol]{Dipartimento di Scienze Fisiche e Sezione INFN, Universit\`a
di Napoli `Federico II', I80126 Napoli, Italy.}
\thanks[pres-ganil]{Corresponding author.\\
Permanent address: GANIL, B.P.~5027, F-14076 Caen Cedex, France.\\
E-mail: frankland@ganil.fr. Tel.: 33~231~454628. Fax: 33~231~454665}
\thanks[cada]{present address: DRFC/STEP, CEA/Cadarache, F-13018
Saint-Paul-lez-Durance Cedex, France.}
\begin{abstract}

A sample of `single-source' events, compatible with the multifragmentation of
very heavy fused systems, are isolated among
well-measured \gdu\ reactions by examining the evolution of
the kinematics of fragments with $Z\geq 5$ as a function of the dissipated
energy and loss of memory of the entrance channel.
Single-source events are found to be the result of very central collisions.
Such central collisions may also lead to
multiple fragment emission due to the decay of excited
projectile- and target-like nuclei and so-called `neck' emission,
and for this reason the isolation of single-source events is very difficult.
Event-selection criteria based on centrality of collisions, or
on the isotropy of the emitted fragments in each event, are found to be 
inefficient to separate the two mechanisms,
unless they take into account the redistribution of fragments' kinetic energies
into directions perpendicular to the beam axis.
The selected events are good candidates to look for bulk effects
in the multifragmentation process.

\end{abstract}



\end{frontmatter}

\section{Introduction}

The study of the behaviour of nuclear matter under extreme conditions of
temperature and pressure, and the possible associated phase transitions,
constitutes one of the major axes of research in nuclear physics
today. The privileged experimental tools to conduct this study are
heavy ion collisions from the sub-relativistic to the ultra-relativistic
energy range, coupled with powerful large-acceptance
multidetector arrays.
With beam energies of a few tens to a few hundreds of MeV per nucleon one
hopes to explore the environs of the
`liquid-gas' coexistence region, which is predicted to be located at
less-than-normal densities
and at temperatures below the critical
temperature for the phase transition ($T_C\approx
10$--16MeV)~\cite{jaqaman}.
In such collisions excited pieces of nuclear matter may be formed with
excitation energies comparable to or even greater than nuclear
binding energies. 

In this bombarding energy range, which has been explored at various experimental
facilities around the world over the last decade (e.g. GSI Darmstadt,
NSCL Michigan, GANIL Caen), an evolution is observed from evaporative
processes, in which light particle emission is the principal mode of
decay of hot nuclei with temperatures $\lesssim 5$MeV, to the so-called
multifragmentation regime where large numbers of nuclei with a wide
variety of masses are observed in the exit channel of individual
collisions~\cite{moretto,guerreau}. The goal of multifragmentation
studies is to link this experimental observation to the properties of
the nuclear matter phase diagram, which entails disentangling dynamical
effects linked to the collision phase 
from decay products of thermodynamically equilibrated excited nuclear systems.

For heavy-ion collisions between 20 and 100~\mpn\ incident energy, the
binary character of almost all reactions has been established
experimentally~\cite{bbmfr,blott,binary1,binary2,binary3,vmetiv}
either by the identification of surviving projectile-like and target-like
fragments (PLF and TLF, respectively) or through the observation of
reaction products' characteristics compatible with their origin in the
statistical decay of two principal primary excited nuclei,
the quasi-projectile (QP) and quasi-target (QT).
It was first thought that such binary reaction mechanisms were essentially a
continuation of the deep-inelastic collisions well-known at lower
energies~\cite{lefort,huizenga}, accompanied by increasingly important
``preequilibrium'' or ``prompt'' emissions due to the opening of the
phase-space for nucleon-nucleon collisions.
However, increasingly exclusive
experiments with 4$\pi$ detector arrays have shown that
an important fraction of the detected particles and heavier fragments
originate from the rapidity region between the projectile and
target~\cite{stuttge},
and that this contribution is difficult to distinguish from that
which may be ascribed to the decay of fully equilibrated
QP and QT~\cite{jerzy,thomasdiane}.
They may, depending on the species considered, the beam energy, system
size and impact parameter, find their origin in e.g. prompt emissions from the
overlap (participant) zone between the two colliding
nuclei~\cite{eudes,pawlow}, decay
of highly-deformed QP/QT~\cite{stefanini,jcolin}, or the formation and rupture of a
necklike structure between projectile and
target~\cite{montoya,toke-neck,pb+auneck,larochelle}.

Faced with such a highly complex situation for most of the collisions,
the so-called ``single-source" events observed for a
small ($\sim 1\%$) part of the cross-section for collisions of very
heavy ions around the Fermi energy~\cite{pb+aumethod,dag96,nm} are of
great interest. In these events, all of the emitted fragments and
particles, apart from a small preequilibrium
component of light particles, seem to originate in the multifragmentation
of a single nuclear system containing almost all of the available mass
and energy of the entrance channel, which simplifies enormously the analysis.
Such events provide a
unique opportunity to study the decay of well-defined and very heavy 
pieces of excited
nuclear matter, for which one expects that bulk effects, 
if present (for example the
mechanical
instability associated with spinodal
decomposition~\cite{bertsch,maria94,spinolett}), should play a dominant
role.
Semi-classical transport
calculations~\cite{gdutheorie,jouault} predict these
reactions to occur for central collisions of heavy nuclei
($b<0.3b_{max}$). 
The initial system formed should therefore suffer large amounts of
compression and heating, and may subsequently expand into the low-density
coexistence region.
 
The method by which a certain set of data is selected depends on the
experimental conditions, detector systems and the reaction studied.
In the present work we present in detail the procedure by which single-source
events have been brought to light and isolated 
for  \gdu\ reactions studied with INDRA. The method used (based on the
angle \tf\ signifying the polar deviation from the beam direction
of the events' principal axis
) has
already been published elsewhere~\cite{pb+aumethod,nm}, but no
comparison has been made before between this method and other procedures
which are far more commonly used to sort experimental data in the
specific context of single-source event selection.
This is the goal of this paper.
In the accompanying paper~\cite{gadoue-ii} we will use these data in order to
probe the role of bulk effects in the
multifragmentation of such heavy systems.

The experimental set-up, including the detector array specifications and
operating conditions during the experiment, is presented in
Sec.\ref{experiment}. 
The first step towards an unbiased global reconstruction of the recorded
reactions is to ensure that a very large proportion of the emitted
products have been measured event by event : this is achieved by an
initial selection of `complete events' (Sec.\ref{compev}).
In order to introduce the comparison between
different selection methods, the most commonly used global variables are
presented in Sec.\ref{ips} and \ref{gsv}. Then the evolution of the
reactions from binary collisions to single-source events
is described in Sec.\ref{evselwilcz}
by classifying them according to \tf . Finally in
Sec.\ref{clamet} we show that the most commonly-used selection methods
are generally less efficient for isolating single-source events by comparison
with the \tf\ selection.

\section{Experimental details}\label{experiment}

The \gu\ system was studied with the $4 \pi$ multidetector INDRA, operating 
at the GANIL accelerator. INDRA, which was described in detail in~\cite{pouthas,pouthas_electron}, can be viewed as an ensemble of 336 telescopes covering
\twid\ 90\% of the \fpi\ solid angle. The detection cells are distributed
amongst 17 rings centred on the beam axis. Low energy
identification thresholds and large energy ranges were obtained through the
design of three-layer telescopes, composed of an axial-field ionization
chamber operated at 30 mbars of $C_3F_8$, a 300 $\mu m$ silicon detector and a
CsI(Tl) scintillator, thick enough to stop all emitted particles,
coupled to a phototube. Such a telescope can detect and identify from protons
between 1 and 200 MeV to uranium ions of 4 GeV. Past 45~\deg , where fast
projectile-like fragments are no longer expected, the telescopes comprise
only two stages, the ionization chamber operated at 20 mbars and the
scintillator. Finally the very forward angles (2\upto 3~\deg ) are occupied by
Ne102--Ne115 phoswiches. Charge resolution of one unit was obtained up
to Z=64
for the fragments identified through the $\Delta E-E$ method in the Si--CsI
couple, and up to Z=20 when the $\Delta E$ signal is furnished by the
ionization chamber (above this limit imposed by the energy resolution of
the ionization chambers, extrapolation of the identification to heavier
fragments is assured by calculations based on energy-loss tables, with a
resolution of a few charge units).

\begin{figure}[htbp!] 
\begin{center}
\includegraphics[width=.75\textwidth]{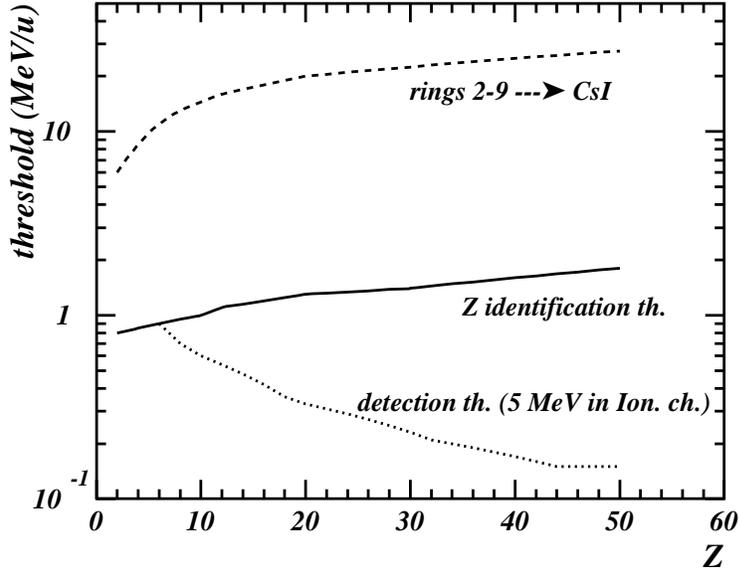}
\caption[THR]{\it Energy thresholds of the 3-member INDRA modules. full
  line: charge identification thresholds; dashed line: minimum energy to
  reach the CsI; dotted line: detection threshold (ions deposing 5 MeV
  in the ionization chamber, but not identified).
\label{THR}} \end{center}\end{figure}

Identification thresholds increase from \twid\ 0.7
\mpn\ for Z=3 to \twid\ 1.7 \mpn\ for Z=50 (Fig.\ref{THR}). Slow heavy ions
at around or just below the threshold energy, stopping in the detector
directly behind the ionization chamber, are partially identified i.e. are
assigned a minimum atomic number. In our analysis such particles are only
included to permit the global characterisation of the least violent 
collisions, when
fission fragments
of the target nucleus
(see Fig.~\ref{chargedens}(a) and accompanying discussion in the text)
are present. 
Energy calibration of all detectors ensured an accuracy of within 5\%\ .
Great attention was paid to pulse-height
defect calibration of the $Si$-300$\mu$m detectors for fragments
$Z\geq 15$~\cite{tabacaru1}. 
For Rings 2--9 residual energy deposited in $CsI$ scintillators were
derived from energy losses in the (preceding) silicon detectors. 
Energy calibrations of $CsI$ scintillators were only performed for
Rings 10--17 ($45\leq\theta_{lab}\leq 176\deg$) and were recently improved
with a better description of quenching which was previously
over-estimated because $\delta$-electrons were
neglected~\cite{parlogetal}: measured kinetic energies of fragments with 
$Z\approx 15$--35 detected beyond 45$^o$ for the \gdu\ system were noticeably
modified as compared to previously published data~\cite{spinolett} (see
accompanying paper~\cite{gadoue-ii}).

A $^{155}$Gd beam accelerated by the GANIL cyclotrons was used to bombard a
100 $\mu g/cm^2$ U target, enclosed between two 20 $\mu g/cm^2$ carbon layers.
In order to ensure a negligible rate of multiple interactions in the
target, the beam intensity was maintained around 3x10$^7$ pps. The low
target thickness was chosen to allow slow fragments to escape the target.
Due to the high detection efficiency of the CsI for $\gamma/e^-$, a
minimum-bias trigger
 based on the multiplicity of fired modules was chosen.
Two settings were used: first M$\geq$4 and then M$\geq$8
in order to minimise the proportion of triggering events which involve the carbon
backings. Only runs using the M$\geq$8 trigger are analysed here.
In the off-line analysis, events having a
multiplicity of correctly identified charged particles inferior to the
experimental trigger condition were rejected for reasons of coherency
(with the exception of Fig.~\ref{compevfig}). 

\section{Event selection methods and procedures}\label{sel_proc}
\subsection{Removal of poorly measured events from the data sample}\label{compev}

As a first step in our event selection procedure, we seek to exclude from
the data sample poorly-measured events, i.e. ones for which a complete
detection of all reaction products has not been achieved. This is necessary
for two reasons. First of all, we want to characterise those rare events for
which a composite system containing almost all of the mass of the colliding
nuclei undergoes multifragmentation. Secondly, it is an essential
prerequisite for the analyses using global variables that we will use later.
In the case of the \gu\ reaction, a way must also be found to eliminate
reactions of the projectile with the carbon target backing.

\begin{figure}[htbp!]
\begin{center}
\includegraphics[width=.75\textwidth]{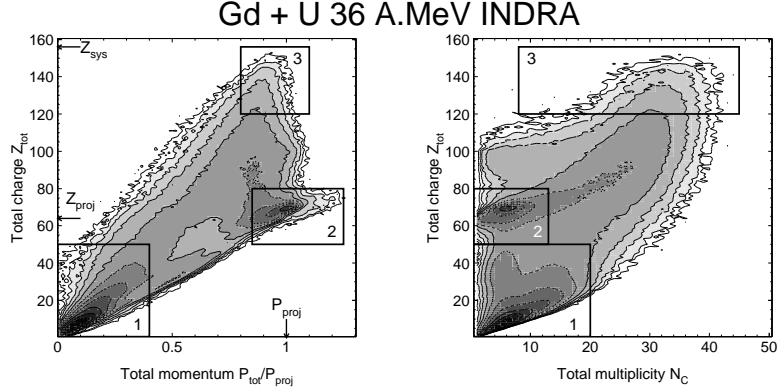}
\caption{\it Non-equidistant logarithmic contour plots showing event-by-event 
correlations between the total detected charge \ztot ,
`momentum' $P_{\mathrm{tot}}$ and
multiplicity of charged products \mutot .
The numbered boxes correspond independently to the three
classes of events discussed in the text.
\label{compevfig}}
\end{center}\end{figure}

The degree of confidence with which we can judge the analysed events to be
`complete' increases with the total detected charge and momentum
(Fig.\ref{compevfig}). Indeed, without making any hypothesis about the
physics of the studied reactions, with a 100\% -perfect detection one
should measure the total system charge
($Z_{sys}=Z_{proj}+Z_{targ}=64+92=156$)
and the beam momentum ($P_{proj}$) in the exit channel for every event.
Fig.\ref{compevfig} shows that experimentally this is far from being
the case.
It should be noted that the `momentum' used here
is calculated from the product of atomic number $Z$
and velocity component in the beam direction $v_z$, and
normalised to the incident (projectile) `momentum' :
\begin{equation}
{P_{tot}\over P_{proj}}
= {\sum Z v_z  \over Z_{proj} v_{proj}}
\end{equation}
This procedure is used to
compensate for the fact that INDRA does not permit isotopic
identification for fragments $Z\geq 4$, and neutrons are not detected.
Simulations of central collisions for the highly neutron-rich \gu\ system
including the effect of secondary particle emission show that this
`charge momentum' is conserved to around
96\% , with a r.m.s. deviation of $\approx$2\% .

The numbered boxes in Fig.\ref{compevfig} correspond to different classes of
detected events. The design of the INDRA detector was optimised for the study
of multifragmentation, typically high-multiplicity ($\lesssim 50$) events
in which reaction products are emitted over a large angular range. On the
other hand, less violent collisions are  not so well detected due to :
\begin{enumerate}
\item identification thresholds (see Sec.\ref{experiment}) which exclude heavy,
slow-moving target residues because of incomplete identification;\\
and
\item loss of rapid projectile-like fragments at laboratory angles smaller
than 2~\deg .
\end{enumerate}

Box 1 events mainly correspond to
rather peripheral collisions for which neither projectile-
(QP) nor target-like (QT) heavy fragments were detected, only light charged
particles (LCP, $Z=1,2$) and occasionally fission fragments from the
target, the latter having very low energies which only permit an approximate
identification via a minimum atomic number $Z_{min}$ (see Sec.\ref{experiment}).
Most of the events in box 2 have total charge $Z_{tot}\approx 70$ and total
momentum close to that of the projectile nuclei. They correspond either to
the detection of a QP and some light particles or to close-to-complete
detection of \nuc{155}{Gd}+\nuc{12}{C} collisions in reverse kinematics.
For total detected charge $Z_{tot}>70$ reactions with the carbon target
support are completely excluded. For measured
$Z_{tot}$ values upwards of 50\%\
of the total charge of the system \gu , increasing `charge-completeness'
basically leads to a global increase of the number of detected products
without much change to the type of reactions observed (except for
low-multiplicity events where the principal detected nuclei are a
quasi-projectile and the target fission fragments).

\begin{figure}[htbp!]
\begin{center}
\includegraphics[width=.75\textwidth]{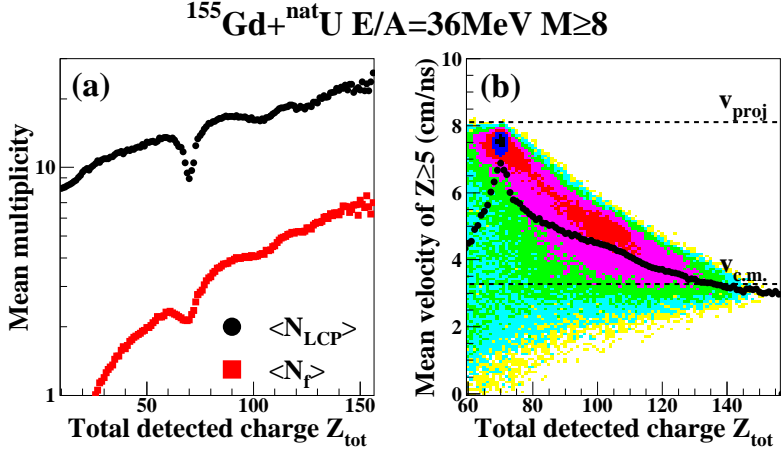}
\caption{\it Effects of the total detected charge $Z_{tot}$ of analysed events.
(a) mean multiplicities of light charged particles ($N_{LCP}$)
and of fragments with $Z\geq 5$ ($N_f$) as a function of $Z_{tot}$;
(b) weighted mean velocity of the $Z\geq 5$
fragments (see text) versus $Z_{tot}$.
Logarithmic intensity scale representing measured cross section.
Points are the mean velocities averaged over each $Z_{tot}$ bin.
\label{effetztot}}
\end{center}\end{figure}

 Because of a lower
overall efficiency for fragment detection compared to LCP, and trivially because of
fragments'
larger atomic numbers, the requirement of highly charge-complete events
slightly favours events where a larger proportion of the available charge is
detected in the form of fragments at the expense of light charged particles
(see Fig.\ref{effetztot}(a) and Figs.\ref{ipsfig}(e),(f)).
This is especially true for fragments emitted backwards in the c.m. frame,
because of the increase in effective thresholds at large laboratory
angles due to the recoil of the c.m.~.
Thus the weighted mean velocity of fragments $\sum m_i v_i / \sum m_i$
is much larger than the \gu\ c.m. velocity $v_{c.m.}$ for small total
detected charge, and indeed close to that of the projectile, $v_{proj}$
(see Fig.\ref{effetztot}(b)): we saw above that in these events all the
backwards-emitted (i.e. target) fragments were missed. As $Z_{tot}$
increases the detected fragments' mean velocity decreases as
more and more backwards-emitted fragments are included.
In other words, those reaction products which have a low
probability of detection are not present in `incomplete' events,
whereas
in `complete' events nearly all emitted
fragments must have been detected, including those emitted into
experimentally unfavourable regions of phase-space, and so the
selection of complete events appears to favour the detection of those products.

In order to retain a reasonably-sized data sample of very well measured
events, we imposed the condition $Z_{tot}\geq 120 (\approx 0.77Z_{sys})$ in the off-line analysis.
We judged this charge-completeness to be sufficient for two reasons. Firstly,
for $Z_{tot}\geq120$ the weighted mean velocity of detected 
fragments (see Fig.~\ref{effetztot}(b)) has an average value close to (within 10\% )
the c.m. velocity of the collisions $v_{c.m.}$, which is a necessary condition
for all of these fragments to have been produced in the disassembly
of a fused system containing most of the mass of the projectile and target.
Angular distributions of fragments and LCP, event shapes and orientations
(see following section) are identical for this data sample whether
calculated in the reaction (\gu ) c.m. frame or in the reconstructed c.m. frame.
Secondly, if only events with $Z_{tot}\geq 0.9Z_{sys}$ are considered the
only significant difference is that the mean fragment multiplicity increases
by $\approx 1$, while the size of the data sample is drastically reduced.
Angular distributions and kinematical observables are unchanged.
Therefore we consider that events with $Z_{tot}\geq 120$ are
on the average sufficiently well-measured to be classified as `complete
events' (box 3 in Fig.~\ref{compevfig}). The additional constraint on the total
detected momentum, $0.8\leq
P_{tot}/P_{proj}\leq 1.1$, has little effect for these events.

The measured cross-section (using target thickness and incident ion
flux) for complete events is 93~mb,
to be compared with a calculated reaction cross-section
value of $\sigma_R =6.5$~barns from systematics~\cite{wilcke}.
An experimental measurement of $\sigma_R$ was
not possible because of the carbon target support.
If this selection seems somewhat draconian, let us recall that it is an
essential condition in order to be able to correctly reconstruct the
kinematics of events where a very large part of the total system has
undergone multifragmentation. We will show in the following that the
accuracy of this reconstruction is of paramount importance for the isolation
of such events.

\subsection{Impact parameter selectors (IPS)}\label{ips}

How may one isolate a sample of events corresponding to the formation and
multifragmentation of a single excited nuclear system ?
A first answer may be to reason in terms of impact parameter. The
`fusion' events that we are looking for must correspond to central
collisions : in a low energy picture for reasons of angular momentum;
 in a high energy picture to maximise the participant zone.

Events may be classed into impact parameter bins using global variables which are
supposed to increase (or decrease) monotonically with $b$ and the geometrical
prescription~\cite{geom} :
\begin{equation}
b_{\mathrm{est}}\left( \Phi_{\mathrm{1}}\right) =
{b_{\mathrm{max}} \over \sqrt{N_{\mathrm{ev}}}}
\sqrt{\int_{\Phi_{\mathrm{1}}}^{\Phi_{\mathrm{max}}} \frac{\mathrm{d}
N}{\mathrm{d} \Phi} \mathrm{d} \Phi }\label{cavata_eqn}
\end{equation}
Here $\Phi$ represents the chosen global variable, in this case reaching its
maximum value $\Phi_{max}$ when $b\rightarrow 0$. $N_{ev}$ is the total
number of recorded events (corresponding to the total geometrical
cross-section $\pi b_{max}^2$), and $b_{est}$ is the estimated impact
parameter for events characterised by the value $\Phi = \Phi_1$. The
assumption underlying Eq.(~\ref{cavata_eqn}) is that one can assign a single
impact parameter to each value of the global variable and vice-versa, i.e.
that fluctuations are negligible.

\begin{figure}[htbp!]
\begin{center}
\includegraphics[width=.75\textwidth]{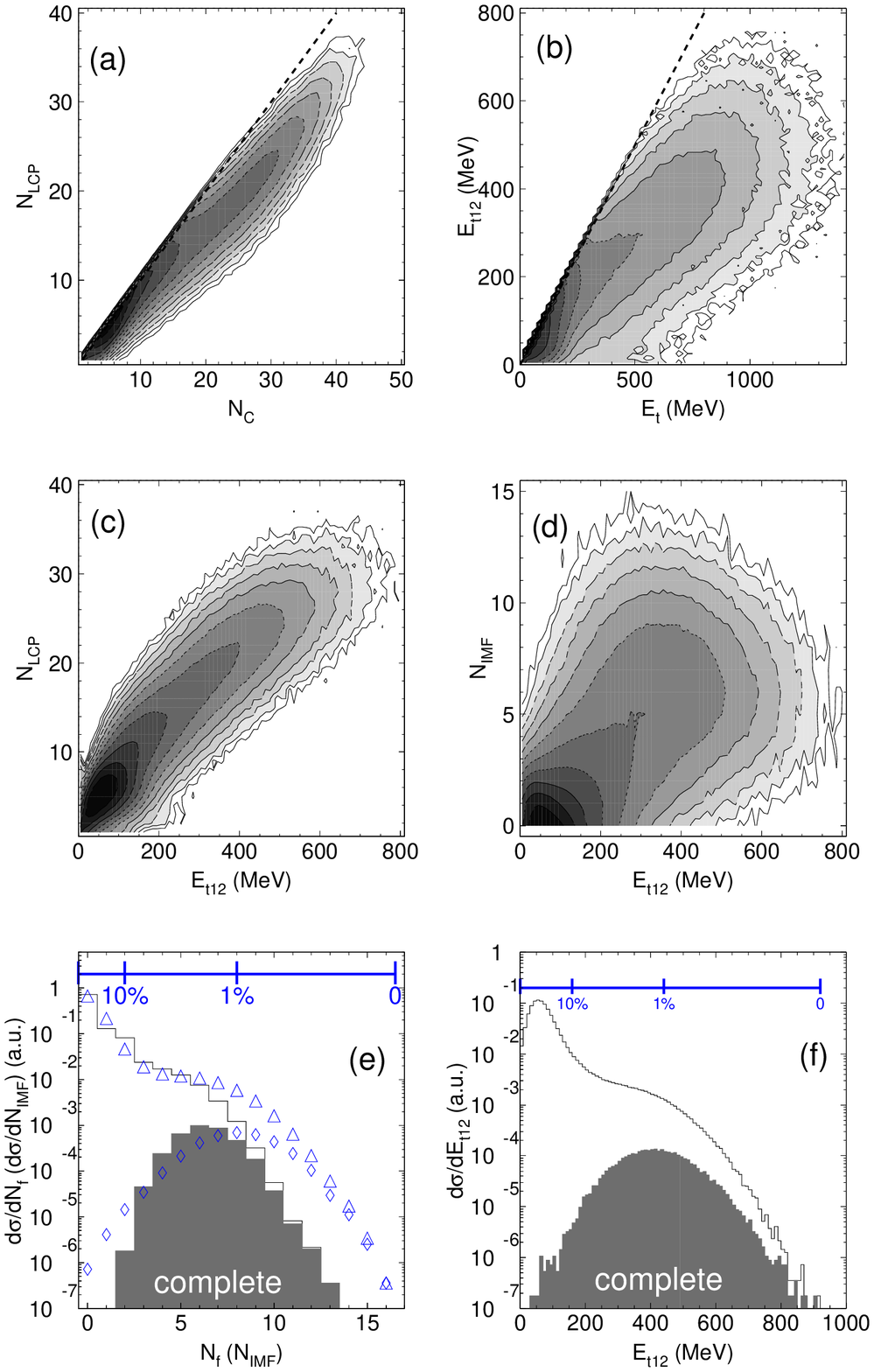}
\caption{\it (a)--(d) Non-equidistant contour plots showing event-by-event 
correlations between \mlcp , \mutot , \etranlcp , \etran\ and \mimf\ (see
text for definitions). (e) Distributions of \nf\ and \mimf\ for all recorded
events (open histogram and triangles, respectively) and for complete events
(shaded histogram and diamonds, respectively). (f) Distributions of
\etranlcp\ for all recorded
events (open histogram) and for complete events
(shaded histogram). An approximate scale in
integrated cross-section, beginning with the `most central collisions'
(largest values of each IPS), is shown in percent for the two variables
\mimf\ and \etranlcp .
\label{ipsfig}}
\end{center}\end{figure}

The most often-used IPS variables are multiplicities (total multiplicity
\mutot , LCP multiplicity \mlcp ) and total transverse energies (of all
particles \etran , or uniquely LCP \etranlcp ). In addition we have defined
the multiplicity of `intermediate mass fragments' (IMF) to be the number
of nuclei with $3\leq Z\leq 30$ (\mimf ), and the multiplicity of fragments
\nf\ for products with $Z\geq 5$. Panels (a)--(d) of 
Fig.\ref{ipsfig} present event-by-event correlations between these
variables.
One may first note the large spread of values around the mean
correlation in each case, therefore fluctuations cannot be neglected, as has
often been observed for dissipative collisions in and below the Fermi energy
range.
Each IPS will thus share
events differently between estimated impact parameter bins.
On the other hand the effect of fluctuations on Eq.(~\ref{cavata_eqn}) is to 
underestimate the impact parameter for the 10\%\ most central
collisions~\cite{geom,peter,jdf_these}.
On Fig.~\ref{ipsfig} approximative integrated cross-sections
are shown for \mimf\ and \etranlcp\ which are calculated with respect to the
total number of recorded events, irrespective of $Z_{tot}$. As this number
includes reactions between the beam and the carbon target backings (see
Sec.~\ref{compev}) the percentages given are a lower limit.
Complete events therefore belong to the $\geq 10$\%\
most central \gu\ collisions measured
(Fig.\ref{ipsfig}(e) \&\ (f)) and further `impact parameter selection'
e.g. multiplicity cuts may well be ineffectual.

Let us mention in passing the variable $Z_{bound}$ used by the Aladin
collaboration~\cite{aladin,stuttge}. It corresponds to the sum of atomic numbers of all fragments
with $Z\geq 2$. In collisions between 100\mpn\ and 1AGeV using the Aladin
spectrometer $Z_{bound}$ represents the size of the projectile spectator
minus evaporated hydrogen isotopes. In this energy range the near-geometrical
dependance of spectator size on impact parameter makes $Z_{bound}$ a good IPS.
The variable $Z_{bound}$ 
does not behave as an IPS with the complete detection of
particles and fragments from both target and projectile.
In another context, it can be useful for model comparisons if it is defined
to be the total charge of all fragments (e.g. with $Z\geq 5$) when comparing data
with calculations whose principal aim is to reproduce fragment partitions
but not all the LCP emitted at different stages of the reaction (see
accompanying paper~\cite{gadoue-ii}). For complete events of \gu\
the average value of the total
charge contained in fragments $Z\geq 5$ corresponds to approximately 50\% of the
total system charge. 

\subsection{Global shape variables (GSV)}
\label{gsv}

A more direct method of discriminating between
different reaction mechanisms is based on
considering how fragments
are distributed in the centre of mass
momentum or velocity space on an event-by-event basis (the `event
shape')~\cite{llope}.
We carry out this analysis using fragments with $Z\geq 5$ rather than the more
usual $Z\geq 3$ definition because of the very heavy nature of the system (note
that both definitions are quite arbitrary).
LCP are left out of the analysis as they result from several
different mechanisms not directly related to fragment production (e.g.
pre-equilibrium emission,
evaporation from hot fragments) and also their smaller masses can blur event
shapes in velocity space.
Then the formation of `fused systems' containing most of the incident
nucleons is characterised by isotropic emission of fragments
in the centre of mass frame, assuming for simplicity complete
relaxation of the multifragmenting system's form i.e. a spherical source,
and negligible angular momentum (spin). On the other hand
the existence of two principal moving sources of emission (e.g. the 
two partners of a deeply-inelastic
collision), or of spectator-like fragments separating from some `participant
zone' around mid-rapidity, implies a rod-like elongated event shape. 
The variables used to exploit this difference in event shape come from
high-energy particle physics, where pattern recognition in momentum distributions 
has long been a major
concern~\cite{barlow}. 

\begin{figure}[htbp!]
\begin{center}
\includegraphics[width=.75\textwidth]{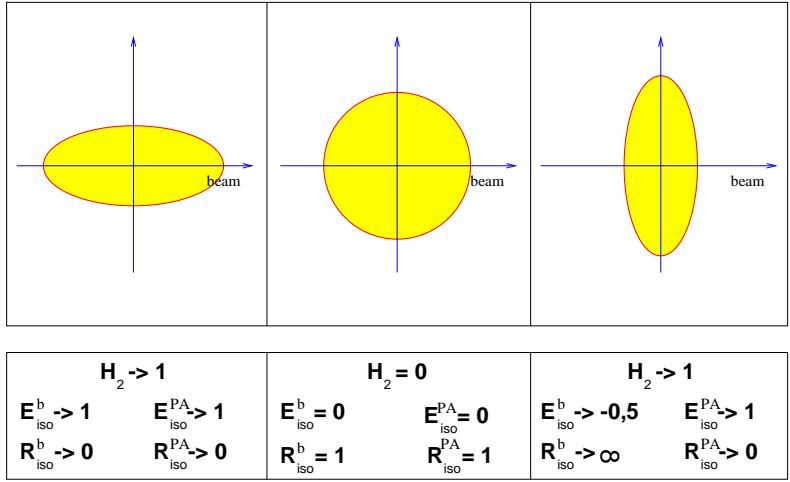}
\caption{\it Values of global shape variables (GSV) for different
event-shapes (represented by an ellipse) 
and orientations with respect to the beam axis. `b' and `P.A.' refer to variables calculated
with respect to the beam axis or the principal axis of the event,
respectively. It should be noted that spherical event shapes (central panel)
can only be achieved for very high (infinite) multiplicities. 
\label{shapes}}
\end{center}\end{figure}

From the cartesian
components of fragment ($Z\geq 5$)
momenta in the centre of mass one may construct the tensor~\cite{tensor},
\begin{equation}\label{tens}
Q_{\mathrm{ij}} \equiv \sum_{Z\geq 5} {p_\mathrm{i} p_\mathrm{j} \over 2m }
\end{equation}
whose eigenvectors and eigenvalues
may be interpreted in terms of an ellipsoid in momentum space, and which give
information on both the event
shape (sphericity, coplanarity etc.)
and its orientation with respect to the beam (see Fig.\ref{shapes}, upper panels).
If $\{\lambda_1,\lambda_2,\lambda_3\}$ are the eigenvalues of
Eq.(\ref{tens}), normalised to their sum and labelled
according to increasing size, $\lambda_3\geq \lambda_2\geq \lambda_1$, then
\begin{eqnarray}
\hbox{sphericity } S & = & \frac{3}{2}\left( 1-\lambda_3\right)\\
\hbox{coplanarity } C & = & \frac{\surd 3}{2}\left( \lambda_2 - \lambda_1\right)
\end{eqnarray}
The `flow axis' corresponding to the largest eigenvalue
$\lambda_3$, or major axis of the ellipsoid,
defines the principal axis of the event, whose deviation from the
beam is referred to by the polar angle \tf . 
This angle varies between 0$^o$ and a maximum value of 90$^o$
because, by convention, it is taken as the angle 
between the forward lobe of the ellipsoid and the beam.
For events which have lost all
memory of the entrance channel kinematics (e.g. single-source events with
negligible angular momentum) this angle is isotropically distributed,
whatever the fragment multiplicity. In the case of a single source with
large angular momentum all angles are populated but directions close to the
beam are favoured~\cite{jdf_these}, due to the preferential emission of
fragments in the reaction plane.

Independently of the tensor one may construct other shape-dependent
variables, for example :
the second moment of Fox and Wolfram, \fox\ \cite{fox}
\begin{eqnarray}
\fox & = & {H(2)\over H(0)} \\
& = & {1\over H(0)} \sum_{1,2}\frac{1}{2} \left| \vec p_1 \right| \,
\left| \vec p_2 \right| \, 
\left( 3 \cos^2 \theta_{rel} - 1 \right) ,\\ 
H(0) & = & \sum_{1,2}\frac{1}{2} \left| \vec p_1 \right| \,
\left| \vec p_2 \right| \label{foxdef}
\end{eqnarray}
where the sums are over pairs of fragments, $\vec p_1, \vec p_2$ are their c.m.
momenta, and $\theta_{rel}$ is their relative angle; the momentum isotropy ratio, \riso\ \cite{keram}, 
\begin{equation}
\riso  =  \frac{2}{\pi}\left( \sum \left| \vec p_{\perp} \right| /
\sum \left| \vec p_{\parallel} \right| \right) \label{risodef}
\end{equation}
where the sums are over all fragments, and $\vec p_{\perp}, \vec p_{\parallel}$
represent perpendicular and parallel projections of $\vec p$; and an energy isotropy ratio 
such as
\eiso\ \cite{erat1},
\begin{eqnarray}
\eiso & = & 1-\frac{3}{2} \left( E_\perp /
\sum_i E_i \right)\\
E_\perp & = & \sum_i E_i \sin^2 \theta_i
\label{eisodef}
\end{eqnarray}
where the sums are again over fragments, $E_i$ is the c.m. kinetic energy of the
$i\/^{\mathrm{th}}$ fragment, and $\theta_i$ its polar angle. An alternative
choice, with a slightly different definition, is the variable
$ERAT$ used by the FOPI collaboration~\cite{erat2}.

By construction \fox\ is independent of the choice of 
basis axes in the c.m. momentum space (it only depends on relative
angles between pairs of fragments), whatever the event shape.
The values of \eiso\ and \riso\ for spherical events are also
independent of the choice of axes. However, spherical events require 
infinite multiplicities of fragments even if the latter are isotropically
emitted, and so are never observed~\cite{daniel}.
This is because the event-shape deduced from a small number
of isotropically-emitted fragments is only a poor approximation
to the `true' spherical distribution,
due to the extremely restricted sampling of the available momentum
space event by event.
As Fig.\ref{shapes} shows, because the event shape is always non-spherical
\eiso\ and \riso\ have to be calculated with respect to the principal axis
of the event in order to ensure non-ambiguous shape determination.
When calculated with respect to the beam axis their values depend not only
on the event shape but also its orientation in the velocity
space, \tf .
In the following we will implicitly suppose that isotropy ratios are
calculated unambiguously, unless explicitly stated otherwise.

The effectiveness of event-shape discrimination is weakened
by the fact that we cannot unambiguously define what
values of the GSV should correspond to the formation and decay of
fused systems because for small multiplicities the event-shape depends
strongly on the number of emitted fragments. On the other hand, an
isotropic distribution of flow angles for a set of events is an unambiguous
signal which is independent of the fragment multiplicity that fragments
have lost all memory of the entrance channel. This fact will be exploited in the following section in order to select single-source events.

\begin{figure}[htbp!]
\begin{center}
\includegraphics[width=.75\textwidth]{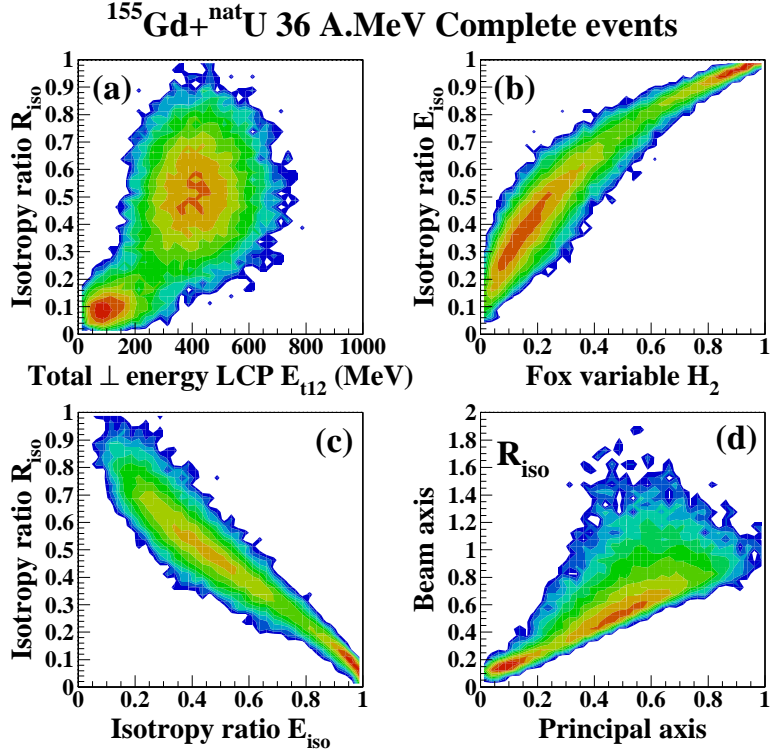}
\caption{\it Event-by-event correlations, for complete events,
between: (a) event-shape (\riso ) and impact parameter selector \etranlcp ;
(b) \eiso\ and \fox ; (c) \riso\ and \eiso (isotropy ratios calculated
with respect to the principal axis of each event);
(d) \riso\ calculated with respect to the beam axis and with respect to
the principal axis of each event.\label{gsvfig}}
\end{center}\end{figure}

Fig.\ref{gsvfig} presents some experimental correlations of these global
shape variables for complete events (an event-shape analysis has no
sense unless performed for complete events).
Fig.~\ref{gsvfig}(a) shows the
correlation between a global shape variable, \riso , and an impact
parameter selector, \etranlcp .
Two regions are apparent: one corresponding to very elongated events
and quite large impact parameters ($\riso <0.25$, $\etranlcp < 300$~MeV);
the other covering a very broad range of both event shapes and transverse
energies.
The former set of events are the least dissipative collisions which are sufficiently violent in order to be well-measured i.e. which are not excluded by the requirement of complete events. We will see in the next section that the principal products of these reactions are residues of the quasi-projectile and of the target fission fragments which are well-separated in velocity space.
As for the remaining events, they correspond to strongly dissipative collisions
and it should be noted that the event shape and impact parameter observables
show no strong correlation.
We will show the existence of `fusion' i.e. single-source events
for a fraction of these collisions in the following section.

Figs.\ref{gsvfig}(b) and (c) show the experimental
correlations which exist between
the three GSV. All of them well separate the two classes of events remarked
in Fig.\ref{gsvfig}(a), especially \fox .
\eiso\ shows a greater variation than \fox\ for the most dissipative collisions, which may mean that it has a greater sensitivity to variations among very
compact event-shapes. On the other hand the two isotropy ratios,
calculated from fragment momenta or kinetic energies,
are very strongly correlated and may be considered to give equivalent information on event shapes.

Finally in Fig.\ref{gsvfig}(d) we present the correlation between
the values of the isotropy ratio \riso\ calculated with respect to the 
beam axis (`ambiguous' shape variable) or the principal axis of the event
(`unambiguous' shape variable).
For the most elongated event shapes (small \riso )
the two methods of calculating \riso\ give very similar values, because the
least dissipative events have mean \tf\ angles of $\approx$10\deg\
(see Fig.\ref{wplot}), therefore the principal axis and the beam axis are
almost aligned.
The majority of the more compact events 
(principal axis isotropy ratio $>0.25$)
also have axis-independent isotropy ratios
(crest following a line ``beam axis \riso =principal axis \riso''),
and this is once again because \tf\ remains quite small for these events
(Zone 2 of Fig.\ref{wplot}).

However there are also events for whom the isotropy ratio
(i.e. the apparent event shape)
depends strongly on the axis chosen, for example appearing as ``spherical''
($\riso =1$) with respect to the beam axis but non-spherical ($\riso <1$)
with respect to the principal axis of the event.
Note also the existence of events for whom the isotropy ratio with respect
to the beam axis $\riso >1$,
which implies an event shape elongated perpendicular to the beam direction.
In both of these cases the `ambiguous' apparent event shape is due to the
increase of \tf ,
which modifies the projection of the fragment momenta on to the beam axis.
For a given unambiguous \riso\ approaching 1, the truly increasingly spherically-symmetric
events populate `beam axis event shapes' which vary less and less with \tf ,
as one would expect. However
we do not observe an `island' of truly spherical events with $R_{iso}^{beam}=R_{iso}^{P.A.}=1$. Once again this is because for the fragment multiplicities
in play, a spherical event shape can not be attained due to finite number effects,
even if all fragments are emitted isotropically.

\section{Event classification using the `Wilczy\'nski diagram': isolation of single-source events}\label{evselwilcz}

\begin{figure}[htbp!]
\begin{center}
\includegraphics[width=.75\textwidth]{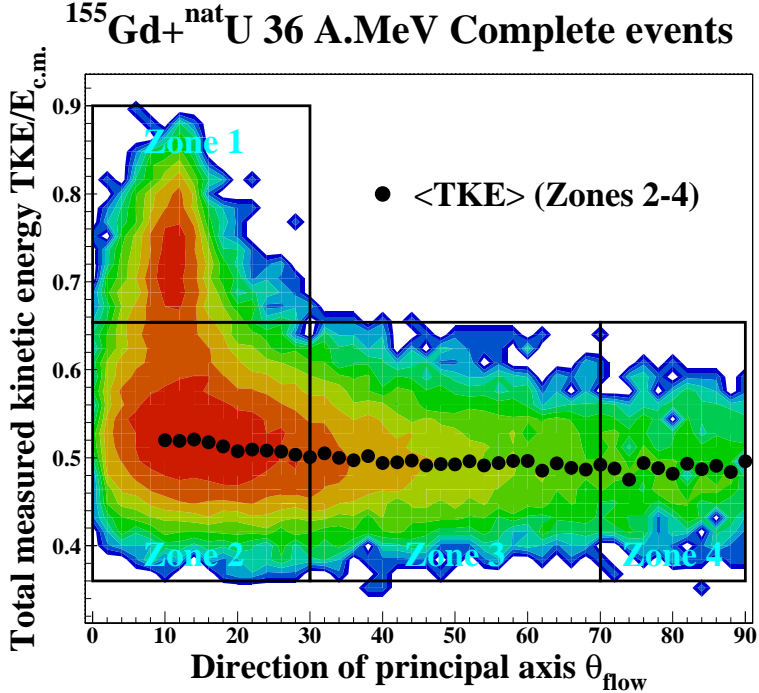}
\caption{\it `Wilczy\'nski diagram' for complete events:
logarithmic intensity scale representing measured cross-section
as a function of total
measured c.m. kinetic energy (as a fraction of the available centre of mass
energy) and `flow' angle or direction of the event principal axis, \tf .
The four zones indicated are used to classify complete events (see text).
For events in Zones 2 to 4 the mean value of TKE is indicated (points) for each
\tf\ bin.
\label{wplot}}
\end{center}\end{figure}

In order to
classify \gdu\ collisions we use a method which was first
employed for the analysis of Pb+Au 29~\mpn\ reactions studied with the multidetector
Nautilus~\cite{pb+aumethod}. 
The correlation between total measured c.m. kinetic energy of detected
charged products (TKE) and the principal
direction of fragment `flow' (\tf ) when plotted for each event
(see Fig.~\ref{wplot})
resembles the Wilczy\'nski diagram~\cite{wilczynski}
well-known at bombarding energies below 20 \mpn\  .
TKE provides a dissipation scale for the collisions, decreasing as more
and more excitation energy is deposited in the system.
For complete events we may write
\begin{equation}
\hbox{TKE }=E_{c.m.}\, +\, Q\, -\, \sum\, E_{neutron}\, -\, \sum\, E_{\gamma}  \label{tke_eqn}
\end{equation}
where $E_{c.m.}$, $Q$, $\Sigma E_{neutron}$ and $\Sigma E_{\gamma}$ are, respectively,
the available centre-of-mass energy, the mass balance of the reaction and the
total neutron and gamma ray kinetic energies. The increase of \enex\ with
increasingly dissipative collisions implies higher
average multiplicities and energies
of neutrons, light charged particles, and
fragments : the latter account for an increasingly negative mass balance.

\begin{figure}[hbtp!]
\begin{center}
\includegraphics[width=.75\textwidth]{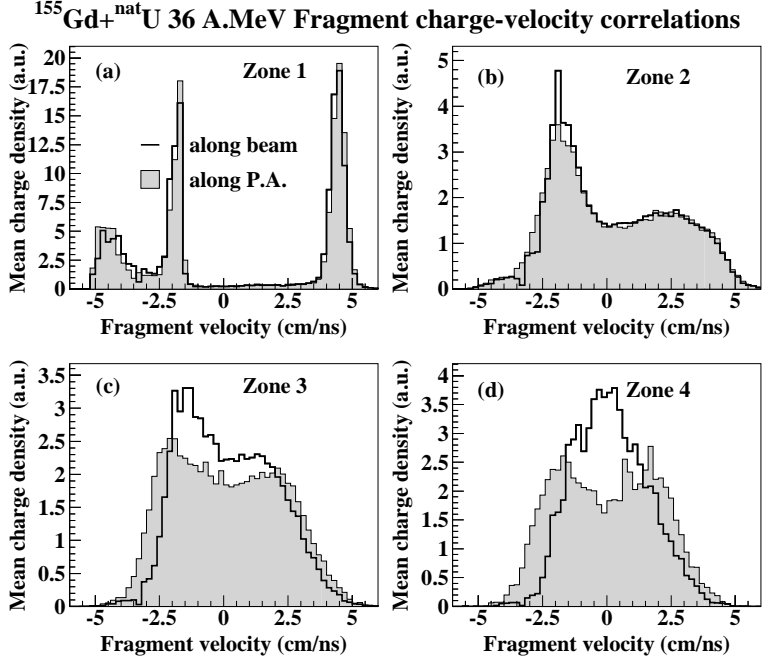}
\caption{\it Fragment charge--velocity correlations
for the four classes of events corresponding to the zones defined in the
Wilczy\'nski diagram (Fig.\ref{wplot}).
The mean charge density (see text) is plotted as a function of
the velocity parallel to the beam axis (thick lines) or the 
principal axis, P.A., of each event derived from the tensor Eq.\ref{tens} (shaded histograms).
\label{chargedens}}
\end{center}\end{figure}

Just as at low energy (for heavy systems), a
dog-legged correlation appears
with a crest running from forward-peaked, slightly dissipative 
collisions (large TKE) to highly-damped reactions with little or 
no memory
of the entrance channel (large \tf ).
We have defined four classes of events depending on their position in the
`Wilczy\'nski diagram', Fig.\ref{wplot}.
In order to examine the topology of the fragment emission in velocity space
in each class of events, we present the charge-velocity correlations 
for each Zone, Fig.\ref{chargedens}.
Here, as a function of velocity along the beam axis or the principal
axis of each event, the average charge density of fragments
is reported, which we define as the
average charge of all the fragments with $Z\geq 5$ which fall
in a given bin in parallel velocity, normalised to the width
of the bins. This is a simplified version of the tool first presented
 in~\cite{digitalfilt}.
The distributions of Fig.\ref{chargedens} therefore give an insight
into the repartition in velocity space of the charge bound in fragments.

In Zone 1 are found the least dissipative collisions
which are sufficiently violent in order to be well measured.
The direction of the principal axis for these events
remains close to the grazing angle ($\theta_{gr}^{c.m.} =
9.8\deg$~\cite{wilcke}), and the principal reaction products 
are residues of the quasi-projectile ($v\approx 4.5$~cm/ns)
and of the target fission fragments ($v\approx -2$~cm/ns and -4.5~cm/ns, respectively).
Let us note in passing that although Fig.\ref{chargedens}(a) shows
clearly that the majority of charge bound in fragments is concentrated
in projectile- and target-like products,
the velocity space between projectile and
target (the so-called mid-rapidity region) is not empty
but populated by a few intermediate mass fragments
which may result from the decay of a `neck' of matter
between the two principal partners of the
reaction, as has been observed for mid-central collisions of 
other heavy systems around the Fermi bombarding energy
(see for example~\cite{montoya,toke-neck,pb+auneck,jerzy}).

\begin{figure}[htbp!]
\begin{center}
\includegraphics[width=.75\textwidth]{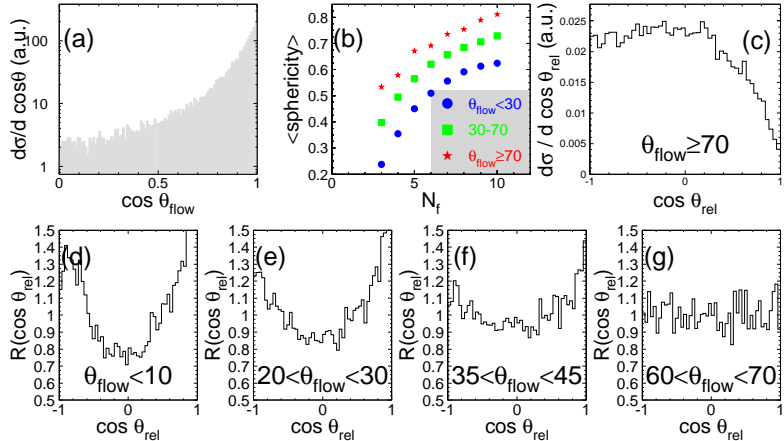}
\caption{\it For the most dissipative collisions (Zones 2--4 of
Fig.\ref{chargedens}): (a) `flow' angle distribution; (b) mean sphericity of the
events as a function of the fragment multiplicity \nf ; (c) distribution of
relative angles $\theta_{rel}$ between pairs of fragments
in Zone 4 events; (d)--(g) evolution of fragment-fragment relative
angle distributions with $\tf$, each distribution has been divided by the 
distribution for Zone 4 events. 
\label{gdusource}}
\end{center}\end{figure}

The most dissipative collisions (Zones 2 to 4) populate all \tf\ angles,
although in majority events are still concentrated at
small angles, $\tf <30\deg$ (Zone 2 in Fig.\ref{wplot}).
The $\cos\tf$-distribution becomes flat (isotropic) for large flow angles
(Fig.\ref{gdusource}(a)).
The repartition of the fragments' charge
is more homogeneous than in Zone 1: specifically, the mid-rapidity
region is fully populated.
However, in Zone 2 (Fig.\ref{chargedens}(b)) there are two asymmetric `bumps' in the distribution
at forward and backward velocities, strongly reminiscent of the entrance channel.
Although not strictly correct, we will refer to this type of 
fragment emission topology as `binary' events in the following.
In Zones 3 and 4 (Fig.\ref{chargedens}(c) and (d)),
due to the increasing deviation of the principal axis from the beam direction, \tf ,
the charge density is not distributed in the same way according to one or the other axis.
Thus, with respect to the beam direction,
the two components' relative velocity becomes smaller
in Zone 3 than in Zone 2 while in Zone 4 only one component
peaked at the c.m. velocity is seen; 
on the other hand, in the velocity-space along the principal axis,
two components are present whatever the Zone and their relative velocity decreases
only very slightly while the distribution becomes
symmetric with regard to the c.m. velocity.

This result is difficult to interpret. 
The distributions with respect to the beam axis suggest
that as \tf\ increases fragments
are produced in the break-up of increasingly relaxed projectile- and target-like
primary fragments, giving way to fusion-like reactions for $\tf \geq 70^o$.
However, this would be in contradiction with
the fact that the mean TKE of events 
in Zones 3 and 4 is constant, and only slightly lower than in Zone 2 (Fig.\ref{wplot},
points). The energy dissipated in the collisions
is practically constant for Zones 2 to 4, which is, on the other hand,
consistent with the velocity distributions with respect to the principal axis
of each event.
One can then understand the apparently spurious
evolution of the two components' relative velocity along the beam direction
to be due to the effect on the projections on to the beam axis
of increasing \tf .

Does this then mean that fragments in Zone 4 result from the break-up of
projectile- and target-like primary fragments, moving apart almost perpendicularly
to the beam ?
This is not so clear, as in fact
both distributions of Fig.\ref{chargedens}(d) are compatible
with the break up of a fused system in to a small number of fragments. 
This is because event-shapes for low-multiplicity reactions are never
spherically symmetric (see Sec.\ref{gsv}), therefore one can always define
event-by-event an oriented principal axis.
In the case of the isotropic break-up of a fused system 
the principal axis has no physical significance and can take all
directions, therefore angular distributions with respect to the beam
are isotropic.
On the other hand the principal axis lies by definition in the direction
of maximum elongation of each event shape, therefore with respect to this
axis fragments seem to be preferentially emitted in the forward and backward
directions, with equal probability~\cite{jdf_these}.
Thus Fig.\ref{chargedens}(d) does not permit to conclude on the origin of the
fragments, and we must examine their characteristics in greater detail.

Mean event-shape sphericity as a function of
fragment multiplicity is shown in Fig.\ref{gdusource}(b) for the three Zones
in \tf\ corresponding to the most dissipative collisions (Zones 2--4). One
should first of all note that for each class of events the increase of sphericity with
fragment multiplicity is expected from the finite number effects
we discussed above.
We see that in general event sphericity increases with `flow' angle, and
that for each fragment multiplicity the most compact events are found in
Zone 4 ($\tf\geq 70\deg$). This is a strong signal of the evolution of
fragment kinematics towards single-source emission, as : (i) the event sphericity and
\tf\ are \emph{a priori\/} independent observables;
and (ii) only comparisons of event shapes for constant multiplicity avoid
distortions due to the aforementioned finite number effects.

One may examine in more detail this evolution by looking at
the distributions of relative angles
($\theta_{rel}$) between
pairs of fragments emitted in the same event.
Fragments emitted by the same source have an isotropic distribution of
$\theta_{rel}$ angles in the rest frame of the emitter, except for
small $\theta_{rel}$ which are suppressed by Coulomb repulsion between
nascent fragments in the case of rapid successive emissions.
Fig.\ref{gdusource}(c) shows exactly this type of distribution for
fragment-fragment relative angles in the Zone 4 events, which
are therefore compatible with the fast break-up of a single source.
Figures~\ref{gdusource}(d)--(g) show the evolution of the
relative angle distributions with the flow angle of the events.
In each case, the distribution
has been divided by that of the Zone 4 events (Fig.\ref{gdusource}(c)),
in order to highlight any differences between the two event samples (a value of
1 for all angles means that the considered distribution is identical to that
for Zone 4 events).
The fact that small and large relative angles between fragment pairs are favoured compared to Zone 4
in events with $\tf < 10\deg$ (Fig.\ref{gdusource}(d))
reveals the dominance of emission from two distinct sources
moving apart in the c.m. frame in these events:
large relative angles are populated by fragment pairs coming from different sources,
while the repopulation of small relative angles is due to kinematical focusing in the direction of the source velocity of fragments 
born of the same source.
The decreasing relative population of large and small angles when
increasing the flow angle of the events 
reflects the diminishing importance of collisions
leading to projectile- and target-like primary fragments.
It should be noted that relative angle
distributions such as that shown in Fig.\ref{gdusource}(f) can also be
obtained if fragments are emitted from a spherical
source having very large angular momentum~\cite{jdf_these}.

To summarise the results presented in this section,
we have shown that the `Wilczy\'nski diagram' (Fig.\ref{wplot})
allows to sort well-measured reactions in to classes of events
(`Zones')
according to the total measured kinetic energy in the c.m. frame TKE
and the flow angle \tf . We have presented the evolution of these reactions
from the least violent collisions, where fission of the target nucleus
is observed, to highly dissipative `binary' collisions, in which many fragments
are produced but they retain a strong kinematical memory of the projectile-target asymmetry. This evolution concerns the first three Zones of the `Wilczy\'nski diagram'.
In Zone 4 are found events for which :
(i) no memory of the entrance
channel remains; (ii) no projectile- or target-like fragments are observed;
(iii) mean event shapes are the most compact of all the very dissipative
collisions;
(iv) fragment kinematics are consistent with rapid emission
from a single source; (v) moreover this single source must contain a very
large proportion of the available mass and energy i.e. preequilibrium
emission is limited (otherwise for an asymmetric system in direct kinematics
one would expect the `fused' system to recoil with a velocity less than that of
the system centre of mass, cf. Fig.~\ref{chargedens}(d))~\cite{fuchs}.

The `Wilczy\'nski diagram' therefore permits the isolation of a sample of
single-source events for $\tf\geq 70\deg$. Let us recall
that for single-source events the flow angle takes all values, therefore
single-source events must also be present for $\tf <70\deg$.
However as we have seen, events with small \tf\ are dominated by a `binary'
fragment emission topology, and keep a strong memory of the entrance channel.
Only at large flow angles do single-source events become dominant and separable
using a \tf\ cut. Of course, this selection method can only function if
the flow angle retains a memory of the direction of the primary
projectile- and target-like fragments coming from highly dissipative
binary reactions, and this direction must remain close to the beam axis. 
It is possible that this may only occur for heavy systems, for which
binary highly dissipative collisions are strongly focused around the grazing
angle due to Coulomb effects~\cite{pb+aumethod,nm}. 
If binary collisions of light systems lead to large flow angles even with low 
cross-sections then single-source
events, when present, may not be revealed by a \tf -cut. Other more refined
selection methods must be applied~\cite{annemarie} in this case.

To conclude this section we will give an estimate for the cross-section
associated with the formation and multifragmentation of fused systems for
\gdu\ reactions. The measured
cross-section for events in Zone 4 is 2.6~mb, which is $\approx$3\%\
of the total cross-section for complete events.
Now we must estimate the fraction of all single-source events represented
by this sample taking into account (i) the \tf\ selection and (ii) the 
selection of complete events.

If all the single-source events have an isotropic flow angle distribution
the fraction of these events that we have selected is given by 
\mbox{$\cos 70\deg =0.34$}. This is the case assuming the formation of
spherical fused systems with negligible angular momentum. In a less ideal scenario
we should allow for deformed and/or turning fused systems formed at finite
impact parameters. The $\cos\tf$-distribution in this case would be
forward-peaked~\cite{arnaud} and the fraction of single-source events
retained by the \tf\ cut would therefore
be less than 34\% . This fraction must however be much
greater than 3\%\ (fraction of all complete events found in Zone 4) because
Zone 2 and Zone 3 events are clearly not dominated by single-source events.
Let us suppose as a conservative estimate that Zone 4 events correspond to
20\%\ of all single-source events. We will use this as an upper estimate for the total cross-section. As a lower estimate we will suppose that the \tf -distribution is indeed isotropic and that the Zone 4 event sample is polluted by
`binary' events. Given the characteristics of these events presented above
we think that an upper estimate for this pollution is around 10\% .

As far as the selection of complete events is concerned, these single-source multifragmentation events
belong to the class of very violent collisions for which the efficiency of
the INDRA detector array is optimal. However the geometrical efficiency of 
90\%\ is perhaps not attained due to the very high multiplicities of these
events (double hits). The distribution of \ztot\ for Zone 4 events is a monotonously decreasing function of \ztot\ (it is of course truncated at $\ztot =120$ because of our complete events selection) and can be fitted by the tail of
a gaussian function of mean value $<\ztot >=115$ and width $\sigma =13$. We estimate the fraction of all single-source events retained by the complete-event selection to
be the fraction of the area under this gaussian curve contained between the
limits $\ztot =120$ and $\ztot =156$ which we found from numerical integration
to be 37\% .

With these hypotheses
we can estimate an upper limit for the total single-source cross-section
(assuming a deformed spinning source, assuming the Zone 4 sample to be free of `pollution' and correcting for event completeness)
to be
\begin{equation}
\sigma_{1source}^{max} = 2.6\mathrm{ mb}\times\frac{1}{20\%}\times\frac{1}{37\%} = 35\mathrm{ mb}\label{sigma_upper}
\end{equation}
while a lower limit (assuming a pollution of Zone 4 by binary events, assuming only spherical zero-spin fused systems and correcting for event completeness)
is given by 
\begin{equation}
\sigma_{1source}^{min} = 90\%\times2.6\mathrm{ mb}\times\frac{1}{34\%}\times\frac{1}{37\%} = 19\mathrm{ mb}
\end{equation}
Let us remark that if one makes the extreme assumption that all of the most
dissipative complete events are compatible with the multifragmentation of
deformed single sources~\cite{boubour} (3\%\ instead of 20\%\ in Eq.~(\ref{sigma_upper}))
then the upper limit for the cross section associated with the multifragmentation of fused systems would be estimated to be 234~mb ($\approx$3\%\ of the total reaction cross section).

\section{Can single-source events be isolated using other selection methods ?}\label{clamet}

We have shown how a sample of single-source events may be isolated based on
the degree of rotation of the event ellipsoid, \tf . 
This sample represents at most 34\%\ of all the single-source events present
among the most dissipative events. One might wonder if another approach,
perhaps based on the more commonly-used tools of impact
parameter selection or event-shape discrimination discussed in Sec.\ref{ips}
and Sec.\ref{gsv}, would not be better adapted to isolate a larger data sample
e.g. by including the small-\tf\ single-source events ?
With this goal in mind we will now look in more detail at the behaviour of
IPS and GSV with regard to the `Wilczy\'nski' event classification scheme.

\begin{figure}[htbp!]
\begin{center}
\includegraphics[width=.75\textwidth]{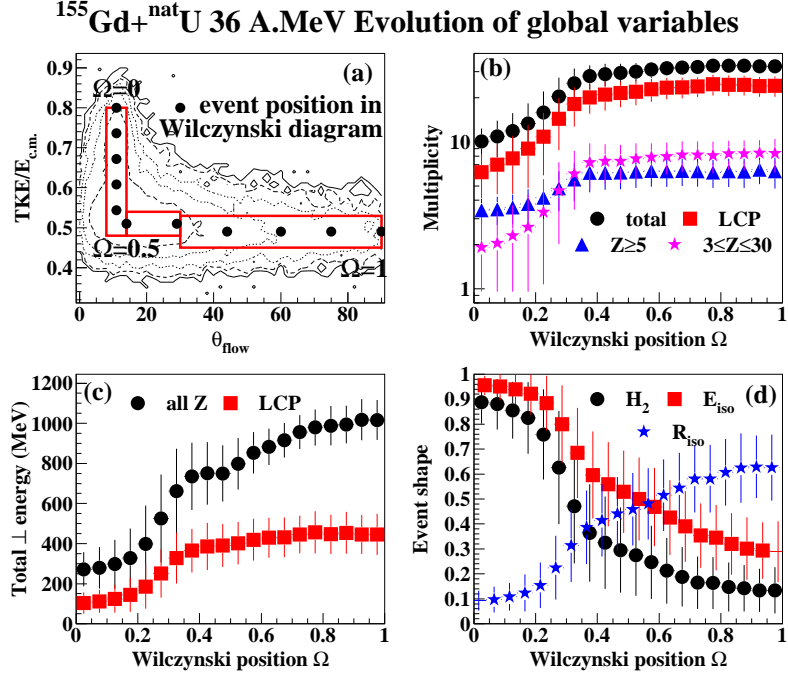}
\caption{\it (a) `Wilczy\'nski diagram' (Fig.\ref{wplot}) for complete events,
showing the definition of the event position variable, $\Omega$. Points
 correspond to $\Omega=0,.1,.2,\ldots$. (b) Evolution of total (\mutot ),
LCP (\mlcp ), $Z\geq 5$ (\nf ) and $3\leq Z\leq 30$ (\mimf )
multiplicities as a function of event position in the `Wilczy\'nski diagram'. (c) Evolution of the total transverse energy of all charged products, \etran , and of LCP, \etranlcp .
(d) Evolution of the event shape according to \fox , \eiso\ and \riso (the two
isotropy ratios are calculated according to the principal axis of each event).
Points correspond to mean values, vertical bars to standard deviations.
\label{bigevol}}
\end{center}\end{figure}

Fig.\ref{bigevol}
presents the evolution of the mean values of the IPS
and GSV variables presented in Secs.~\ref{ips} and~\ref{gsv}
as a function of event position in the `Wilczy\'nski diagram'.
In order to do this we defined an observable, $\Omega$, which varies as a function of TKE and \tf\ in such a way that it increases monotonously along the crest of the experimental correlation observed in Fig.\ref{wplot} and
Fig.\ref{bigevol}(a). $\Omega$ is therefore a global variable related to the position of events in the `Wilczy\'nski plot'. Its value is given by
\begin{equation}
\Omega = {0.5\over 0.32}\times\left( 0.8 - {\mathrm{TKE}\over E_{c.m.}}\right)
\end{equation}
for $8\deg\leq\tf\leq 14\deg$ and $0.48\leq\mathrm{TKE}/E_{c.m.}\leq 0.8$,
covering values $0\leq\Omega\leq 0.5$, and by
\begin{equation}
\Omega = {0.5\over 76\deg }\times\left(\tf - 14\deg \right)+0.5
\end{equation}
for $14\deg <\tf\leq 30\deg$ and $0.48\leq\mathrm{TKE}/E_{c.m.}\leq 0.54$
($0.5<\Omega\leq 0.6$) and for $30\deg <\tf\leq 90\deg$ and $0.45\leq\mathrm{TKE}/E_{c.m.}\leq 0.53$ ($0.6<\Omega\leq 1$). Outside of these limits
(represented by rectangular boxes in Fig.\ref{bigevol}(a)) $\Omega$ is undefined and the corresponding event is not included in the calculated mean. Roughly speaking, events from Zone 1 of Fig.\ref{wplot} have $0\leq\Omega\leq 0.25$;
from Zone 2, $0.25 <\Omega\leq 0.6$; from Zone 3, $0.6<\Omega\leq 0.85$;
and from Zone 4, $0.85<\Omega\leq 1$.

It is clear that the majority of the evolution of the
IPS and GSV variables 
takes place for $\Omega <0.4$ (Fig.\ref{bigevol}(b)--(d)) i.e. Zones 1 and 2
of the `Wilczy\'nski diagram', and this evolution is therefore
strongly correlated to the dissipation of the available 
kinetic energy $E_{c.m.}$.
We have shown that these events retain a strong memory of the entrance channel
(Fig.\ref{chargedens}(a),(b) and Fig.\ref{gdusource}(d),(e)). 
IMF and LCP multiplicities
(Fig.\ref{bigevol}(b))
as well as the total transverse energies
(Fig.\ref{bigevol}(c)) all increase as the collisions become more
dissipative,
while event shapes become more and more compact
(Fig.\ref{bigevol}(d))
reflecting the decreasing relative velocity between primary target- and
projectile-like fragments in the exit channel and the increasingly isotropic
emission of fragments in the c.m. frame.

If we now look to the most dissipative collisions
($\Omega >0.4$: Zones 2--4 of Fig.\ref{wplot}) the mean values 
of IPS variables show a slight evolution, much smaller than
the standard deviations of their distributions (vertical bars in Fig.\ref{bigevol}(b)--(d)).
Only \etran\ continues to increase significantly for $\Omega >0.5$
because increasing \tf\ means increasing fragment transverse energies;
however \etran\ is constant for $\Omega >0.8$.
The mean $Z\geq 5$ fragment multiplicity \nf\ is the same
whatever the position of the event in the horizontal branch of the 
`Wilczy\'nski diagram'.
It should be noted that IMF and LCP multiplicities show an
identical saturation with dissipated energy,
unlike what was observed for the
$^{136}$Xe+$^{209}$Bi 28~\mpn\ reaction~\cite{dyna_frag_prod}.

IPS are therefore quite insensitive to the evolution of fragment 
kinematics from `binary' collisions to single-source events that was
shown in Sec.~\ref{evselwilcz}, and they seem to be most strongly
correlated with energy dissipation.
Let us point out that a simple calculation for heavy systems
of the excitation energy per nucleon for a fully-damped pure binary collision (no mass
exchange)\begin{equation}
E^* = E_{c.m.} - \frac{1}{2}{A_p A_t \over A_p+A_t} v_{rel}^2
\end{equation}
(where $v_{rel}$ is given by the Viola systematics~\cite{viola})
or for fusion
\begin{equation}
E^* = E_{c.m.}+ \Delta (A_p,Z_p)+\Delta (A_t,Z_t) - \Delta (A_p+A_t,Z_p+Z_t) 
\end{equation}
(where $\Delta (A,Z)$ is the mass excess of nucleus $^A_ZX$,
using the extrapolation of~\cite{guet} in the case of the fused system)
gives very similar results.
This may explain the lack of significantly different IPS behaviour
for single-source events.

Event-shapes on the other hand continue to evolve towards
more compact forms with increasing \tf\ (Fig.\ref{bigevol}(d),
$\Omega > 0.6$) as in Fig.\ref{gdusource}(b),
but here we can see that, as for IPS, distributions about the mean values are comparatively wide.
It should be noted that although here we mix events
with different fragment multiplicities, the mean $Z\geq 5$
multiplicity is the same for all $\Omega >0.4$ (Fig.\ref{bigevol}(b))
and so the evolution of the GSV
truly reflects a change in event shapes, i.e. the evolution of the fragment
kinematics with increasing \tf .
This evolution is complete
for $\tf\gtrsim 70\deg$ ($\Omega\gtrsim 0.85$):
 it is for this reason that we
choose to define our sample of single-source events (Zone 4 of Fig.\ref{wplot})
as having $\tf\geq 70\deg$
even though the distribution of relative angles between fragments for events
with $60\deg\leq\tf\leq 70\deg$ is identical to that for $\tf\geq 70\deg$ 
(see Fig.\ref{gdusource}(g)).

\subsection{Selecting the most central collisions}
\begin{figure}[htbp!]
\begin{center}
\includegraphics[width=.75\textwidth]{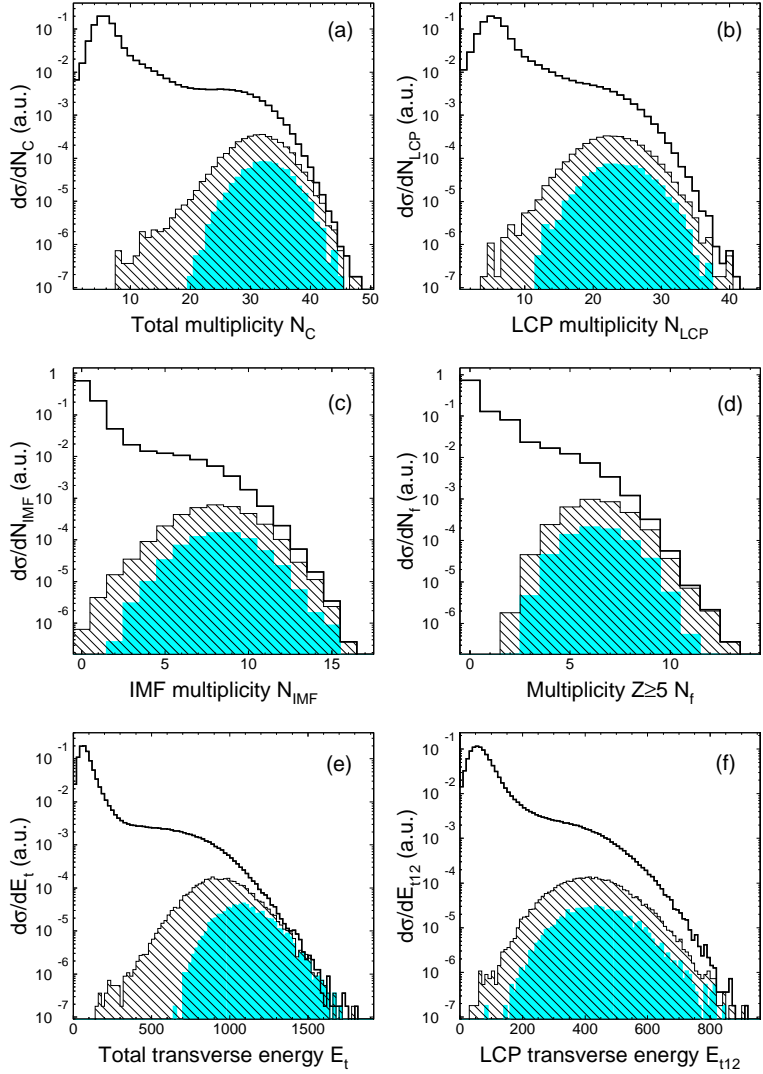}
\caption{\it Comparison of IPS distributions for all recorded events
(histogram), the most dissipative events (Zones 2--4 of Fig.\ref{wplot},
hatched histogram)
and Zone 4 events (shaded histogram). All histograms are
normalised with respect to
the total number of recorded events.
\label{cf_suvar}}
\end{center}\end{figure}

Let us try to define IPS cuts corresponding to very central collisions in
order to isolate single-source events. In Fig.\ref{cf_suvar} the
IPS-variable distributions for Zone 4 events (shaded histograms)
are compared to those for all recorded events (including \gc\ collisions, lines)
and for the most dissipative events (Zones 2 to 4 of Fig.\ref{wplot}, hatched histograms).
All of these variables increase monotonously with
decreasing impact parameter, i.e the largest values correspond to the most
central collisions (see Sec.\ref{ips}). Fig.\ref{cf_suvar} shows
that Zone 4 events correspond to central collisions (large IPS values)
but even though they have mean values of IPS which are higher than for
Zones 2 and 3 (Fig.\ref{bigevol}(b,c)),
the distribution is so wide that the same values are explored 
by Zone 4 events as by all of the most dissipative events.
This observation can be interpreted in terms of a large overlap of
impact parameters leading to single-source or to `binary' exit channels,
with single-source events merely a subset of the most central collisions.
It is clear that no cut on an IPS variable can be defined which would
isolate some or all of Zone 4 events without mixing them with the other Zones. 
The only exception is \etran\ (Fig.\ref{cf_suvar}(e)) for which,
because large \tf\ angles imply large fragment transverse energies,
Zone 4 events are the only ones to explore the largest \etran\ values.

\begin{figure}[htbp!]
\begin{center}
\includegraphics[width=.75\textwidth]{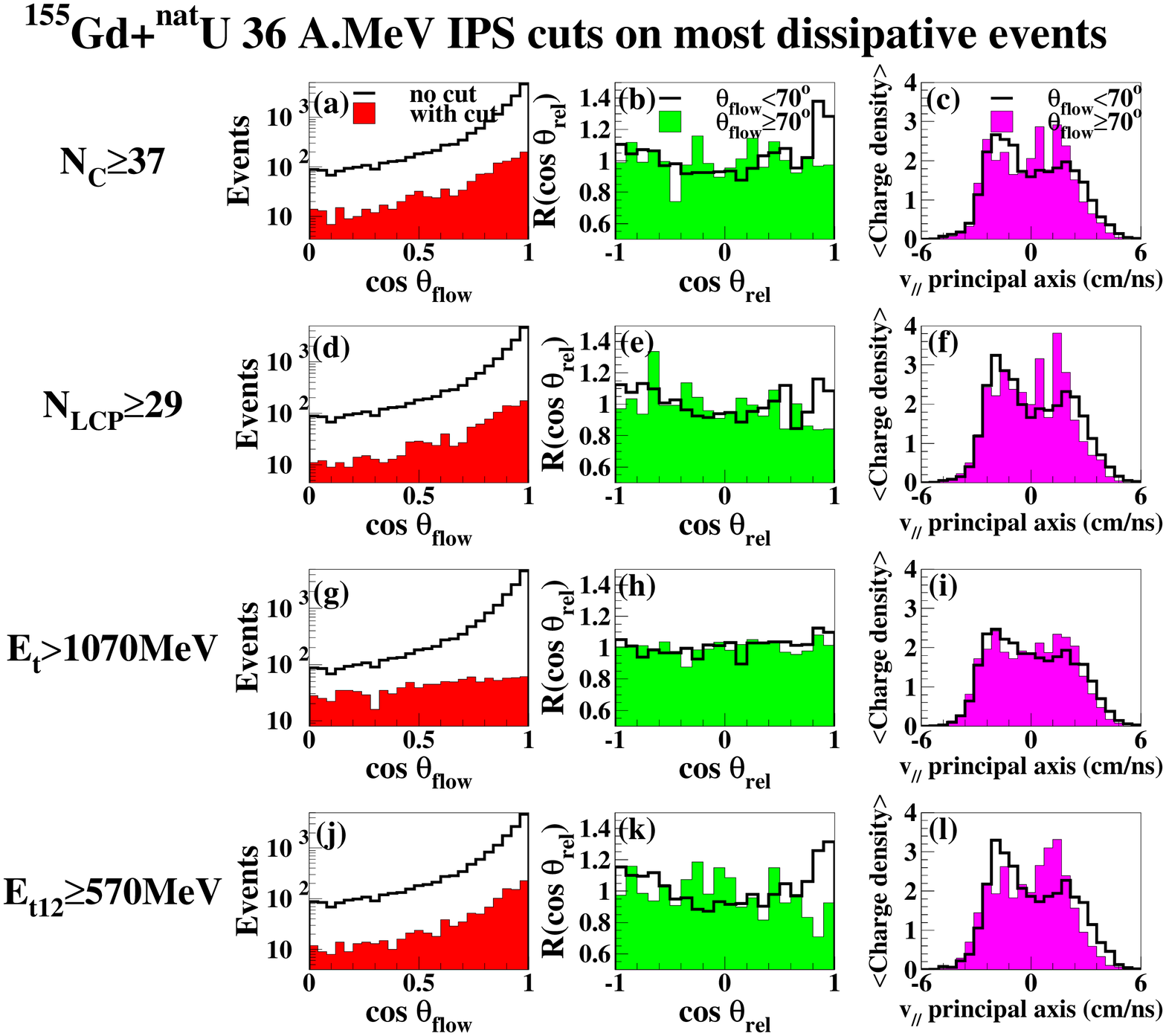}
\caption{\it Results of applying IPS cuts defined in the text to
the most dissipative collisions (Zones 2--4 of Fig.\ref{wplot}). (a),(d),(g),(j): flow angle distribution of the events selected by the IPS cut (shaded histograms) compared to the total distribution for Zones 2--4 (thick lines).
(b),(e),(h),(k): distribution of fragment-fragment relative angles for IPS-selected events divided
bin-by-bin by the distribution obtained for Zone 4 events (Fig.\ref{gdusource}(c)) i.e. $R(\cos\theta_{rel})=1$ for identical distributions.
(c),(f),(i),(l): fragment charge-velocity correlations for IPS-selected events with respect to the principal axis of each event.
In the latter two cases, the events selected by IPS cut are also
separated into those with $\tf <70\deg$ (thick lines) and those with $\tf\geq 70\deg$ (shaded histograms).
\label{cutsips}}
\end{center}\end{figure}

One may then wonder if high-multiplicity or
high transverse energy cuts can be used in order to isolate a sample of
single-source events different to that constituted by Zone 4 of
the `Wilczy\'nski diagram'.
Such cuts have been applied to the most dissipative events i.e. Zones 2 to 4
 of Fig.\ref{wplot} (we have shown that IPS are capable of
 distinguishing between these events and those of Zone 1, Fig.\ref{gsvfig}(a))
and the results are presented in Fig.\ref{cutsips}.
The cuts are : $\mutot\geq 37$, $\mlcp\geq 29$, $\etran >1070$MeV,
$\etranlcp\geq 570$MeV. These values are slightly higher than the mean
values observed for Zone 4 events (cf. Fig.\ref{cf_suvar}), and they were chosen in part so
that the number of events retained by each cut was the same within $10\%$.

Flow angle distributions for the events selected with
\mutot , \mlcp\ and \etranlcp\ cuts
(Fig.\ref{cutsips}(a),(d),(j), shaded histograms)
are forward-peaked and appear to be of the same form as the total distribution
(thick lines) except that the proportion of events with the smallest \tf\ is 
slightly reduced. We see that only $\sim$10\%\ of the Zone 4 event sample
is retained by these IPS cuts.
It should be noted that for $^{197}$Au+$^{197}$Au 35~\mpn\ reactions (very similar in both total mass and available energy to the studied reactions), a much smaller 
total multiplicity cut ($\mutot >24$) was found to be sufficient in order to isolate a sample of highly spherical events with an
isotropic flow angle distribution corresponding to 10\%\ of the
total measured reaction cross-section~\cite{dag96}.
Our results show that such a cut is completely insufficient
for reactions of an otherwise identical asymmetric reaction. 
The cut applied to total transverse energies (Fig.\ref{cutsips}(g)) on the other hand
greatly reduces the proportion of events with $\cos\tf >0.5$
and the resulting \tf\ distribution is nearly isotropic.
The proportion of Zone 4 events retained by this cut
is larger ($\sim$30\% ) than for the other IPS
as is to be expected from comparison of
Fig.\ref{cf_suvar}(e) with Fig.\ref{cf_suvar}(a)--(d) and (f).

If any of these cuts isolates a sample of single-source events,
the characteristics of the events should be independent of the flow angle.
Therefore we separate the events selected with each IPS cut
into two lots, those with small flow angles ($\tf <70\deg$) and those with large
flow angles  ($\tf\geq 70\deg$). It should be recalled
 that the second lot is a subset of the Zone 4 event sample.

Distributions of relative angles between fragments are shown in 
Figs.\ref{cutsips}(b), (e), (h) and (k). In each case, the distribution
has been divided by that of the Zone 4 events (Fig.\ref{gdusource}(c)),
in order to highlight any differences between the two event samples (a value of
1 for all angles means that the considered distribution is identical to that
for Zone 4 events).
For the multiplicity and LCP total transverse energy cuts we see that in the
sub-sample of events with $\tf <70\deg$, large and small fragment-fragment
angles are favoured compared to Zone 4 events (thick lines)
suggesting fragment emission from two sources. This means that the
 \tf\ distributions for events selected by
 these cuts are forward-peaked because they mix single-source and `binary' events.
On the other hand the events selected by the \etran -cut show fragment-fragment
relative angle distributions which are identical to Zone 4 events independently
of the subsequent division according to flow angle.

The average charge density of fragments in the different sets of events are
presented as a function of their velocity along the principal axis in
Figs.\ref{cutsips}(c), (f), (i) and (l). The $\tf <70\deg$ subset of the events selected
by cuts on \mutot , \mlcp\ and \etranlcp\ shows a memory of the entrance
channel through the presence of larger fragments at velocities
$v_\parallel <0$.
Note also that these cuts seem to favour,
among the events with $\tf\geq 70\deg$, events having a larger mean charge
density for $v_\parallel >0$ (heavier fragments emitted in the forward direction).
The \etran -selected events present very homogeneous,
symmetric and \tf -independent charge-velocity correlations, although a
slight increase of the mean charge density is perhaps visible at backwards
velocities for $\tf <70\deg$ (Fig.\ref{cutsips}(i), thick line).

To sum up this section on the selection of the most central collisions,
we have shown that the sample of single-source events
found in Zone 4 of the `Wilczy\'nski diagram' is a subset of the most dissipative events as far as IPS are concerned, and are therefore probably a subset of the most central collisions.
Attempts to isolate single-source events with an IPS cut supposed to select the most central collisions only result in mixed samples containing single-source and `binary' events. The exception to this rule is the total transverse
energy which selects a sample of single-source events with an almost
isotropic flow angle distribution,
albeit with some pollution by `binary' events being in evidence for the smallest \tf . 

\subsection{Selecting the most compact event shapes}
\begin{figure}[htbp!]
\begin{center}
\includegraphics[width=.75\textwidth]{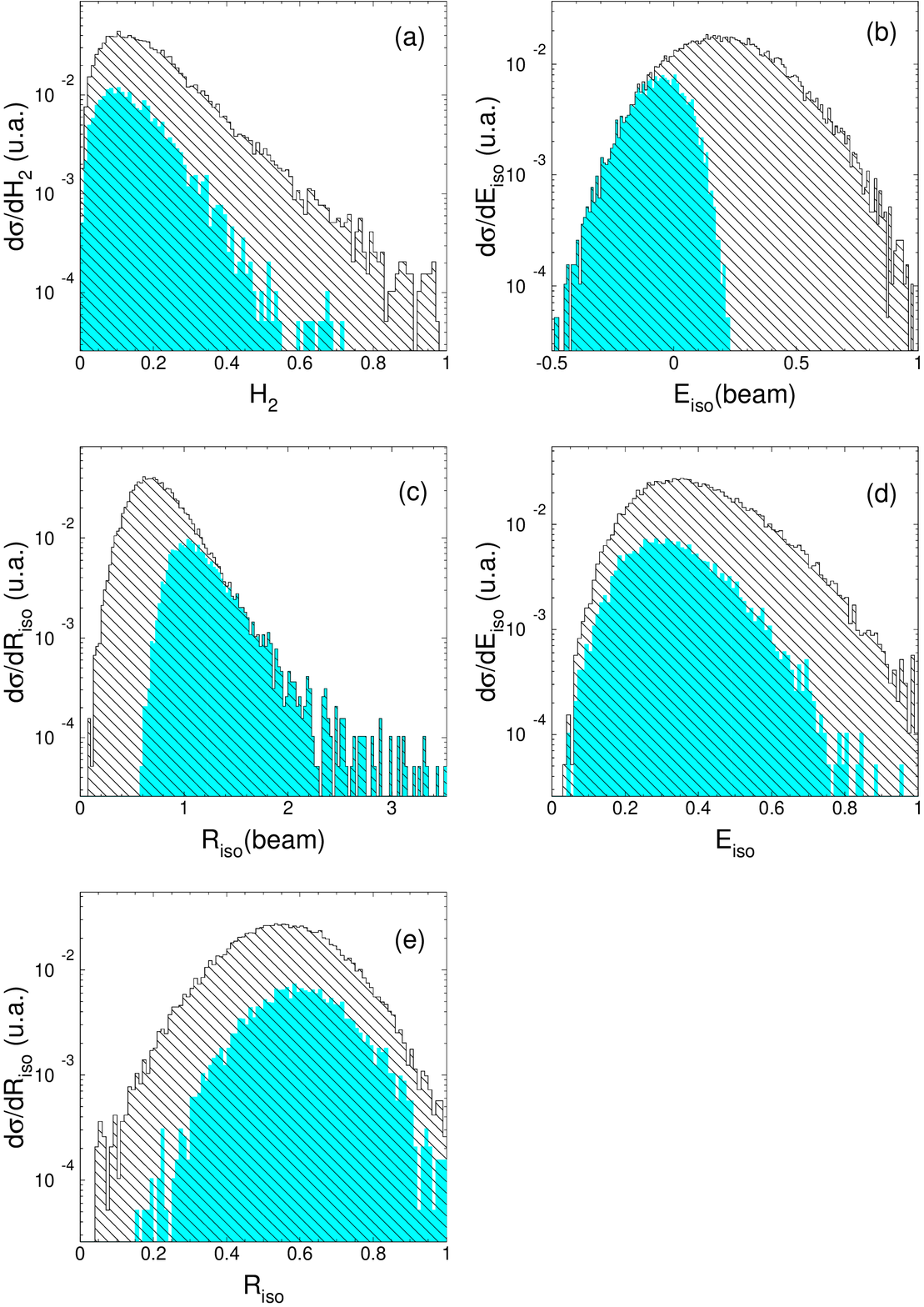}
\caption{\it Comparison of GSV distributions for
the most dissipative events (hatched histogram) and single-source events
($\tf\geq 70\deg$, shaded histogram). \eiso (beam) and \riso (beam) are the
values calculated with respect to the beam axis (cf. $\eiso^b$ and $\riso^b$
of Fig.\ref{shapes}), \eiso\ and \riso\ are calculated with respect to the event
principal axis (cf. $\eiso^{P.A.}$ and $\riso^{P.A.}$ of Fig.\ref{shapes}). All histograms are
normalised with respect to
the total number of complete events.
\label{cf_suvar2}}
\end{center}\end{figure}

Let us now consider the selection of the most spherical (or compact) event shapes,
by defining cuts using global shape variables. Fig.\ref{cf_suvar2} compares the event shapes
of Zone 4 events with those of all the most dissipative events. For all GSV 
(except \eiso\ calculated with respect to the beam axis)
most of the values populated by all of the most dissipative events can also
be associated with Zone 4 events. In all cases it can be seen that the Zone 4
single-source event sample explores a very wide range of event shapes. One may
then worry that cuts made to restrict to only the most compact event shapes will select
a very particular subset of single source events (see below). As in the IPS case,
it is clear that no cut on  a GSV variable can be defined which would isolate all or
some of the Zone 4 events without mixing them with the other Zones, except for the
two `ambiguous' isotropy ratios. For these GSV Zone 4 events are the only ones to explore
values signifying a large proportion of fragment emission in directions perpendicular
to the beam (see Sec\ref{gsv}, Fig.\ref{gsvfig}(d)), because these events have the largest
\tf\ angles.

\begin{figure}[htbp!]
\begin{center}
\includegraphics[width=.75\textwidth]{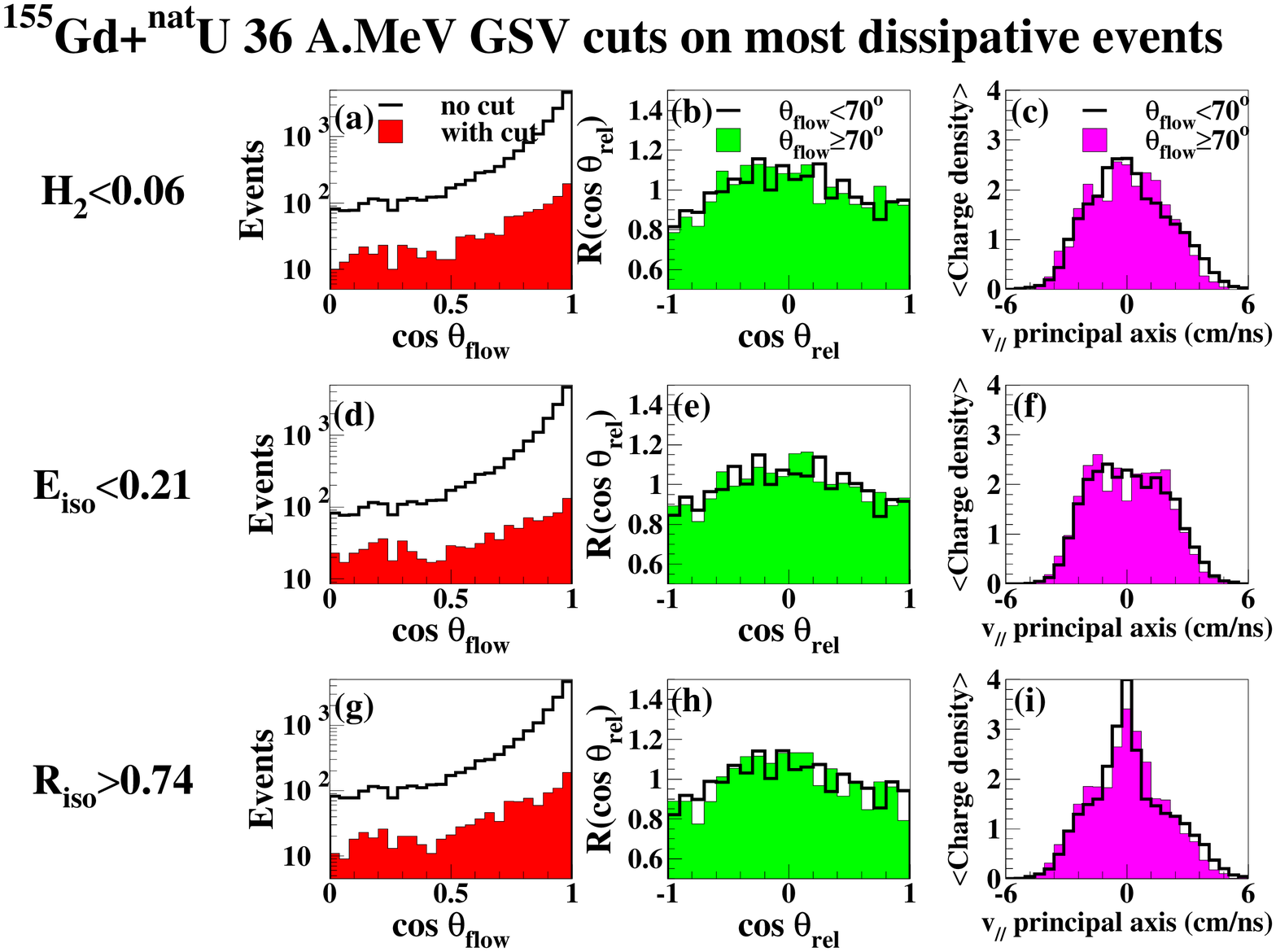}
\caption{\it Results of applying GSV cuts defined in the text to
the most dissipative collisions (Zones 2--4 of Fig.\ref{wplot}). (a),(d),(g),(j): flow angle distribution of the events selected by the GSV cut (shaded histograms) compared to the total distribution for Zones 2--4 (thick lines).
(b),(e),(h),(k): distribution of fragment-fragment relative angles for GSV-selected events divided
bin-by-bin by the distribution obtained for Zone 4 events (Fig.\ref{gdusource}(c)) i.e. $R(\cos\theta_{rel})=1$ for identical distributions.
(c),(f),(i),(l): fragment charge-velocity correlations for GSV-selected events with respect to the principal axis of each event.
In the latter two cases, the events selected by GSV cut are also
separated into those with $\tf <70\deg$ (thick lines) and those with $\tf\geq 70\deg$ (shaded histograms).
\label{cutsgsv}}
\end{center}\end{figure}

We have defined GSV cuts supposed to select the most spherical event
shapes, and in each case the size of the event sample was the same 
(within 10\% ) as those previously selected with IPS cuts.
The cuts were: $\fox <0.06$, $\eiso <0.21$ and $\riso >0.74$ for the `unambiguous' shape variables, $|\eiso |<0.5$ and $0.93<\riso <1.07$ for the isotropy ratios calculated with respect to the beam axis.
First, let us consider the results for the 3 `unambiguous' GSV (Fig.\ref{cutsgsv}).

Selecting the most compact event shapes does not change greatly the form
of the flow angle distributions (Fig.\ref{cutsgsv}(a), (d), (g)).
Events with small \tf\ are still favoured in the selected event samples,
although the \eiso\ cut does slightly flatten the $\cos\tf$ distribution. The fragment-fragment relative angle distributions, 
on the other hand,
are independent of \tf\ for each GSV cut (Fig.\ref{cutsgsv}(b), 
(e), (h)). Moreover the suppression
of small and large $\theta_{rel}$ relative to the distribution for 
Zone 4 events is the opposite of what one would expect for samples
mixing single-source and `binary' events. In fact, these distributions
are directly related to the GSV cuts as can be readily seen by inspection of the \fox\ definition, Eq.\ref{foxdef}. As \fox\ depends on
$\cos^2\theta_{rel}$, imposing small values of \fox\ favours events
where the majority of fragment pairs have $|\cos\theta_{rel}|\approx 0$. Although the link between $\theta_{rel}$ and the isotropy ratios 
is not so trivial to demonstrate analytically, it is very probable that
the same kind of effect is in operation also.

This is not the only non-trivial effect that cuts on `unambiguous' shape variables can have on the event topology. The charge-velocity
correlations show that the \riso\ and (to a lesser degree) \fox\ cuts
favour events in which the heaviest fragments have small velocities
parallel to the principal axis. This may be a way of `compacting'
the event in momentum space.
The charge-velocity correlations for \eiso -selected events on the
other hand are very homogeneous, as one would na\"\i vely expect
for highly compact event shapes.
 
\begin{figure}[htbp!]
\begin{center}
\includegraphics[width=.75\textwidth]{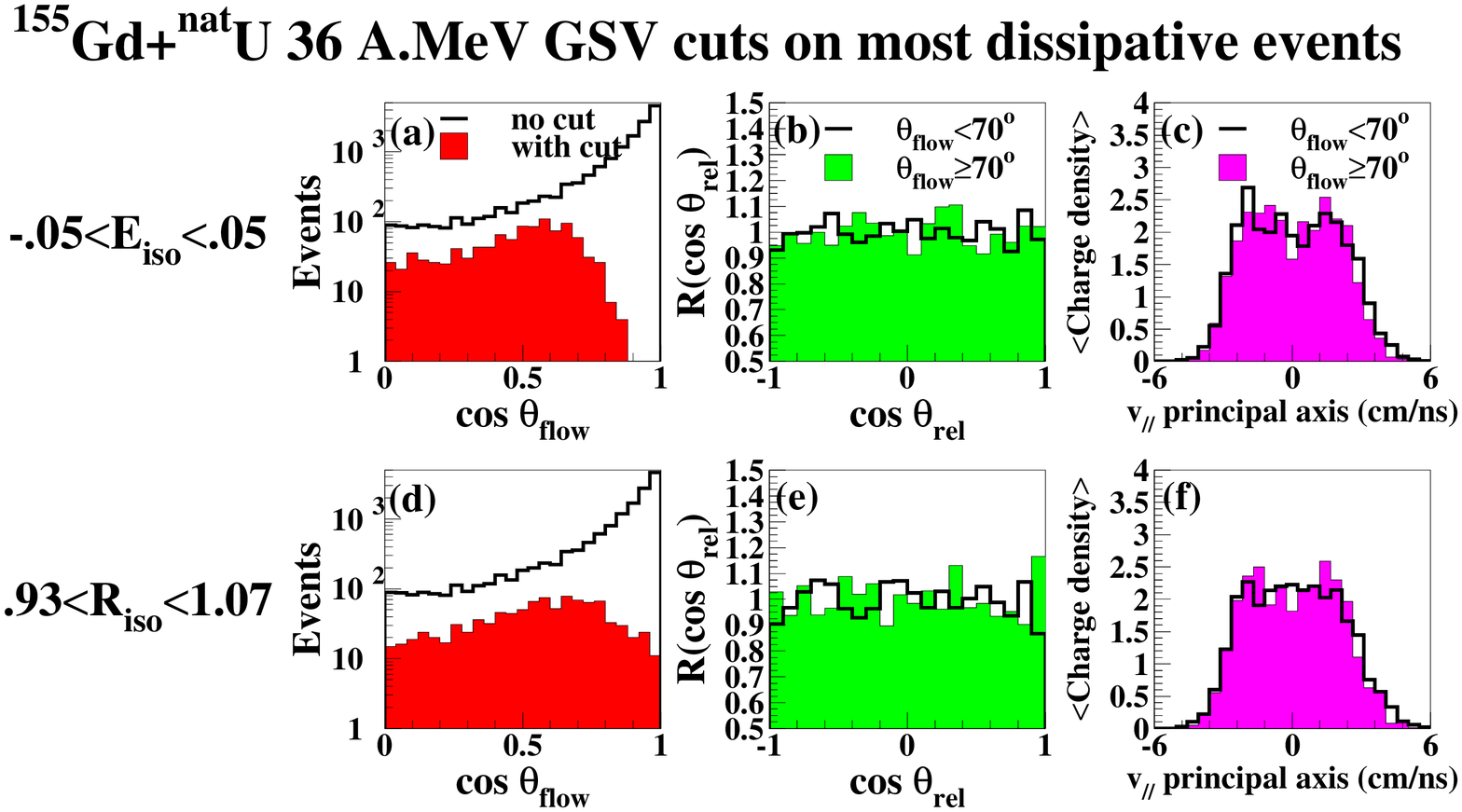}
\caption{\it Results of applying `ambiguous' GSV cuts defined in the text to
the most dissipative collisions (Zones 2--4 of Fig.\ref{wplot}). (a),(d):
flow angle distribution of the events selected by the GSV cut (shaded histograms) compared to the total distribution for Zones 2--4 (thick lines).
(b),(e): distribution of fragment-fragment relative angles for GSV-selected events divided
bin-by-bin by the distribution obtained for single-source events (Fig.\ref{gdusource}(c)) i.e. $R(\cos\theta_{rel})=1$ for identical distributions.
(c),(f): fragment charge-velocity correlations for GSV-selected events with respect to the principal axis of each event.
In the latter two cases, the events selected by GSV cut are also
separated into those with $\tf <70\deg$ (thick lines) and those with $\tf\geq 70\deg$ (shaded histograms)
\label{cutsgsvbis}}
\end{center}\end{figure}

Finally let us see what is the effect of applying to the most dissipative 
collisions cuts designed to select the most compact events using the two `ambiguous' isotropy ratios, \eiso\ and \riso\ calculated with respect to the beam
axis.
The resulting event samples do indeed appear to be good samples of
single-source events: they have fragment-fragment relative angle distributions
which are identical to that of Zone 4 events whatever value takes \tf\ 
(Fig.\ref{cutsgsvbis}(b),(e)) and charge-velocity correlations show an homogeneous, symmetric distribution of the fragments in velocity space
(Fig.\ref{cutsgsvbis}(c),(f)). These cuts therefore avoid the peculiar effects
on event topology that are seen with the `unambiguous' GSV.
However Fig.\ref{cutsgsvbis}(a),(d) shows that the flow angle distribution
of the selected events is unphysical, having been distorted by the correlation
between the `ambiguous' isotropy ratios and \tf . Small \tf\ events are strongly suppressed or even excluded (in the case of \eiso ) by the `ambiguous' GSV 
cuts.

To summarise this section on selecting the most compact events,
we have shown that the single-source event sample isolated in Zone 4 of the
`Wilczy\'nski diagram' explores a very wide range of event
shapes, therefore GSV cuts can only select subsets of single-source events
with very particular fragment emission topologies.
Specifically, the `unambiguous' GSV select events with forward-peaked flow
angle distributions, in which pairs of fragments are preferentially
detected at 90\deg\ and the heaviest fragments are at rest in the centre
of mass frame. On the other hand the isotropy ratios when calculated with respect
to the beam axis permit the extraction of small-\tf\ single-source events
which are similar to the Zone 4 event sample. However most of the single-source
events with small \tf\ ($\cos\tf <0.3$) remain irrecoverable due to the
strong correlation between \tf\ and these variables.

\section{Conclusions}

We have shown for \gdu\ reactions
measured with the INDRA $4\pi$
detector array
how a sample of single-source events may be isolated which correspond to the formation
and multifragmentation of very heavy fused systems, comprising the
majority of the entrance channel nucleons. These events are selected
from among the most well-characterised reactions (at least 80\%\ of the
total charge and momentum were measured) using a condition on the flow angle \tf\
between the principal axis of the event ellipsoid, constructed from
fragment ($Z\geq 5$) kinetic energies, and the beam direction. One
expects events with large \tf\ to show little memory of the colliding
nuclei because of the forward-focused differential 
cross-section for deeply inelastic
collisions of heavy nuclei.
We have shown that a class of events compatible with the fast
break-up of a compact single source is dominant
at large flow angles, $\tf\geq 70\deg$.

This sample of single-source events is associated with the largest
mean multiplicities and total transverse energies of charged reaction products,
and on average the most compact event shapes.
However the single-source events
are only a subset of the most central/most compact events,
therefore isolating a sample of them using impact parameter selectors (IPS)
or global shape variables (GSV) is not as trivial as it may first appear.
We found that the only variables capable of isolating an event sample
with the same characteristics as events with $\tf\geq 70\deg$,
namely \etran\ the total transverse energy of charged reaction products,
\riso\ and \eiso\ the isotropy ratios of momentum and kinetic energy flow,
respectively, calculated with respect to the beam axis,
are all correlated with the transverse kinetic energy of the detected
fragments, and this is also true for \tf .
Therefore using the present methods
it is not possible to obtain an unbiased \tf\ distribution
(which could give important information on the relaxation
in form of the multifragmenting sources, angular momentum, etc.~\cite{arnaud,boubour})
for the totality of the single-source events in the data sample.

The $\tf\geq 70\deg$ cut permits the study of a sample of
very heavy multifragmenting systems, for which bulk properties may play a
decisive role. This work is presented in the accompanying paper~\cite{gadoue-ii}.

\end{document}